\documentclass[]{jfm}

\usepackage{graphicx}
\usepackage{epstopdf,epsfig}
\usepackage{overpic}
\usepackage{newtxtext}
\usepackage{newtxmath}
\usepackage{natbib}
\usepackage{hyperref}
\hypersetup{
    colorlinks = true,
    urlcolor   = blue,
    citecolor  = black,
}

\newcommand{\RomanNumeralCaps}[1]
\linenumbers

\shorttitle{Wave Statistics and Energy Dissipation of Breaking Waves Generated by a Wave Plate}
\shortauthor{S. Liu,  H. Wang, A.C. Bayeul-Lain\'e, C. Li, J. Katz, and O. Coutier-Delgosha}

\title{Wave Statistics and Energy Dissipation of Breaking Waves Generated by a Wave Plate}
\author{Shuo Liu\aff{1} \corresp{\email{shuo.liu@ensam.eu}},
  Hui Wang\aff{1}, {Annie-Claude Bayeul-Lain\'e}\aff{1}, Cheng Li\aff{2}, Joseph Katz\aff{3}, \and Olivier Coutier-Delgosha\aff{1,4}
   \corresp{\email{ocoutier@vt.edu}}}
   
\affiliation{
\aff{1}Univ. Lille, CNRS, ONERA, Arts et Metiers Institute of Technology, Centrale Lille, UMR 9014 - LMFL - Laboratoire de Mécanique des Fluides de Lille - Kamp\'e de F\'eriet, F-59000 Lille, France
\aff{2}Guangdong Technion-Israel Institute of Technology, Shantou, Guangdong, CN
\aff{3}Department of Mechanical Engineering, Johns Hopkins University, 3400 N. Charles Street, Baltimore, MD 21218, USA
\aff{4}Kevin T. Crofton Department of Aerospace and Ocean Engineering, Virginia Tech, Blacksburg, VA 24060, USA
}

\begin{document}
\maketitle

\begin{abstract}
The present study focuses on direct numerical simulations of breaking waves generated by a wave plate at constant water depths. The aim is to quantify the dynamics and kinematics in the breaking process, together with the air entrainment and energy dissipation due to breaking. Good agreement is achieved between numerical and experimental results in terms of free-surface profiles, energy budget, and bubble statistics during wave breaking. A parametric study was conducted to examine the effects of wave properties and initial conditions on breaking features. According to research on the Bond number ($Bo$, the ratio of gravitational to surface tension forces), a larger surface tension produces more significant parasitic capillaries at the forward face of the wave profile and a thicker plunging jet, which causes a delayed breaking time and is tightly correlated with the main cavity size. A close relationship between wave statistics and the initial conditions of the wave plate is discovered, showing that breaker types can be classified according to the ratio of wave height to water depth, $H/d$. Moreover, the energy dissipation rate due to breaking can be related to the initial conditions of the wave plate by applying the conventional dissipation scaling of turbulence theory, and further correlated with the local breaking crest geometry using inertial-scaling arguments. The proposed scaling of the dissipation rate during the active breaking period is found to be in good agreement with our numerical results.
\end{abstract}


\section{Introduction}\label{sec:Introduction}
As a strongly nonlinear intermittent process occurring over a wide range of scales, wave breaking plays an important role in air-sea interactions by limiting the height of surface waves and enhancing the transfer of mass, momentum, and heat between the atmosphere and the ocean \citep{melville1996role,perlin2013breaking}. When a wave breaks, the free surface may experience dramatic changes, entraining air into the ocean and ejecting spray into the atmosphere, with the production of bubbles and aerosols \citep{kiger2012air,veron2015ocean}, and the generation of local turbulence near the free surface. Breaking also controls the fate of oil spills and contaminants in the upper ocean, determines particle size distribution and dynamic transport, and further affects the health of marine environments \citep{delvigne1988natural,deike2017lagrangian, li2017size}. The processes associated with breaking waves have received much research attention, and the greatest progress has been made in the geometry of breaking, breaking onset criteria, dissipation due to breaking, and air entrainment \citep{perlin2013breaking,deike2022mass}.

In particular, the energy transfers involved in waves have been studied extensively over the years, and the parameterization of the dissipation rate due to breaking has benefited greatly from laboratory experiments and numerical measurements. The parameterization originating from seminal experimental studies by \citet{duncan1981experimental} has indicated that the work done by the whitecap or energy dissipation rate per unit length of wave crest scales to the fifth power of a characteristic speed, i.e., $\epsilon_l = b \rho c^5 / g$. Here, $b$ is a dimensionless coefficient related to the wave-breaking strength, $\rho$ is the density of water, $c$ is a characteristic speed associated with the breaking wave, and $g$ is the acceleration due to gravity. The breaking parameter $b$ was first assumed to be a nondimensional constant but subsequently shown by extensive experimental investigations to vary over several orders of magnitude when varying the breaking wave slope $S$ \citep{rapp1990laboratory,tian2010energy}. To establish possible relationships between the breaking parameter $b$ and the initial conditions of breaking waves, the conventional dissipation scaling of turbulence theory has been applied to the wave-breaking process \citep{duncan1981experimental,drazen2008inertial,mostert2020inertial}, following the form of the turbulent dissipation rate based on dimensional analysis \citep{batchelor1953theory}. The local turbulent energy dissipation rate during wave breaking can be estimated as $\epsilon =\chi (w^3/l)$, where $\chi$ is a proportionality constant, $w$ is the representative velocity scale, and $l$ is the turbulent integral length scale characterizing the energy-containing turbulent eddies \citep{taylor1935statistical,vassilicos2015dissipation}. Therefore, the energy dissipation rate per unit length of the crest is $\epsilon_l=\rho A\epsilon$ by assuming a turbulent cloud of cross section $A$. \citet{drazen2008inertial} relates the local turbulent energy dissipation rate to the local breaking properties by inertial scaling, i.e., $\epsilon = \sqrt{gh}^3 / h$, where $h$ is the breaking height and $\sqrt{gh}$ is the ballistic velocity of the plunging breaker. The turbulence cloud is assumed to be a circle with a cross section of $A = \upi h^2 /4$. This indicates that the dissipation rate per unit length of breaking crest $\epsilon_l = \rho A\epsilon \propto \rho g^{3/2} h^{5/2} \propto (hk)^{5/2} \rho c^5 / g$, where $k$ is the wavenumber, and $c = \sqrt{g/k}$ by the dispersion relation in deep water. This leads to $b \propto S^{5/2}$, with $S = hk$ being the breaking wave slope. By considering a breaking threshold, a semiempirical scaling $b=\chi_0 (S-S_0)^{5/2}$ has been introduced. Here, $\chi_0$ is a coefficient, and $S_0$ is the critical slope \citep{romero2012spectral}. This scaling has been extensively confirmed using laboratory experiments \citep{tian2010energy,grare2013growth} and numerical simulations \citep{iafrati2009numerical,deike2016air,de2018breaking}. In addition to deep water breaking waves, the energy dissipated by breaking solitary waves on a beach slope has also been quantified by \citet{mostert2020inertial}. The representative velocity scale is considered to be the impact velocity, which is calculated ballistically as $w=\sqrt{2gH_b}$, where $H_b$ is the wave amplitude at breaking. The turbulent integral length scale is estimated to be the undisturbed depth at breaking $d_b$, and the cross section of the turbulence cloud is assumed to be $A = \upi {H_b}^2 /4$. Therefore, the dissipation rate per unit length of breaking crest is $\epsilon_l = \rho A\epsilon \propto \rho g^{3/2} {H_b}^{7/2} / {d_b} \propto ({H_b} / {d_0})^{7/2}({d_b} / {d_0})^{-1} \rho c^5 / g$, where $d_0$ is the undisturbed depth before the beach slope and $c = \sqrt{g d_0}$ by the dispersion relation in shallow water. These efforts have resulted in relationships between the dynamics and the kinematics of breaking waves, and the parameterization of the dynamics has been developed using geometric properties.

Moreover, breaking is responsible for the air-sea gas exchange through air entrainment. A number of air bubbles may be entrained in the upper ocean during wave breaking, with a wide range of bubble sizes. Small bubbles may be dissolved into water columns, and larger bubbles entrained by breaking may rise back to the surface and collapse \citep{woolf2019key,deike2022mass}. The bubble size distribution $N(r)$ is a key parameter in controlling the mass transfer during the bubble formation process by breaking waves. Laboratory studies have reported measurements of the bubble size distribution beneath a breaking wave using a variety of optical and acoustic techniques \citep{loewen1996bubbles,deane2002scale,leifer2006bubbles,blenkinsopp2010bubble}, while theoretical and numerical investigations have led to a deeper understanding the turbulent bubble cascade above the Hinze scale \citep{chan2021turbulent}.
\citet{garrett2000connection} proposed a dimensional cascade argument to determine the bubble size distribution by assuming that air is injected at a scale much larger than the Hinze scale, and turbulent pressure fluctuations are the dominant mechanism for bubble breakup. Therefore, the bubble size distribution is linearly proportional to the gas input rate per unit volume $Q$ ($Q$ has dimension $T^{-1}$, for dimensional consistency), and the bubble size distribution $N(r)$ (has dimension $L^{-4}$ as it is the number of bubbles per unit radius and per volume) must be of the form $N(r) \propto Q \epsilon^{-1/3} r^{-10/3}$. 
\citet{deane2002scale} presented the results of a detailed study of the bubble size distributions in laboratory and oceanic breaking waves. Depending on the bubble size, two distinct mechanisms controlling the size distribution have been proposed.
The bubble size distribution for bubbles larger than the Hinze scale, determined by turbulent fragmentation, presents a -10/3 power-law scaling with the bubble radius. Bubbles smaller than the Hinze scale are formed by jet and drop impact on the wave face throughout the active phase, with a -3/2 power-law scaling. This bubble size distribution scaling has been confirmed by numerous laboratory experiments using various optical and acoustic techniques \citep{leifer2006bubbles,blenkinsopp2010bubble} and theoretical and numerical investigations \citep{chan2021turbulent,riviere2021sub}.

While great progress has been made in previous studies of breaking-wave dynamics, including the prediction of the geometry, breaking onset, energy dissipation, and air entrainment, some limitations still remain, and important research needs to be done to gain a more thorough understanding of breaking waves. First, breaking waves in deep water have been studied the most. However, breaking waves in shallow and intermediate water depths may experience more dramatic free-surface changes, which makes the problem more complicated. Moreover, consecutive periodic wave trains were usually generated in previous experiments, while further studies of single-wave breaking events are still required to isolate the effect of wave breaking. The direct numerical simulation (DNS) approach, which resolves all breaking processes in waves, has proven to be feasible in deep-water studies \citep{iafrati2011energy,deike2016air} and shallow-water breakers \citep{mostert2020inertial}, but in previous studies, because of limited computational resources, limited wave scales with smaller Reynolds numbers and Bond numbers have been used. Nevertheless, experimental waves with a wide range of length scales, from wave breaking at the metre scale to air bubble entrainment at the micron scale, should be considered.

Thus, in this context, this study focuses on breaking waves produced by a wave plate, emphasizes the early phases of the wave-breaking process defined by \citet{deane2002scale}, and reduces the physics involved to a two-dimensional (2D) issue. Experimental waves comparable to breaking waves in the open ocean are reproduced using DNS under a variety of initial conditions. A wide range of scales have been resolved using an adaptive mesh refinement scheme, retaining a realistic representation of the breaking processes compared with experiments, including the transfer and dissipation of energy and the formation and breakup of bubbles in a two-phase turbulent environment. The paper is organized as follows. In \S \ref{sec:Problem description}, we introduce the configurations of laboratory breaking-wave experiments and a dimensional analysis for waves generated by wave plates. In \S \ref{sec:Numerical study}, we present the numerical scheme and model setup, and conduct mesh convergence analysis and model verification. The wave characteristics with different breaking intensities during wave breaking are analysed in \S \ref{sec:Breaking characteristics}. In \S \ref{sec:Parametric study}, we investigate the scaling of wave dynamics and kinematics to initial conditions by using inertial-scaling arguments and analysing numerical results. We conclude in \S \ref{sec:Concluding remarks} with some summaries of the present work.

\section{Problem description}\label{sec:Problem description}
\subsection{Laboratory breaking-wave experiments}\label{sec:Laboratory experiments}
\begin{figure}
  \centerline{\includegraphics[width=\linewidth]{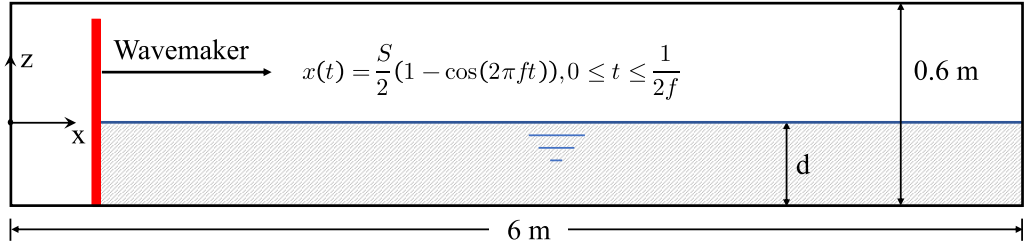}}
  \caption{Sketch of laboratory breaking wave experiment and numerical domain.}
\label{fig:domain}
\end{figure}

This study aims to investigate the wave dynamics, energy budget, and air entrainment during wave breaking, as well as the quantitative relation of the main cavity, breaking criteria, and energy dissipation with respect to the fluid properties and initial conditions, by reproducing experimental waves through direct numerical simulation. A series of breaking-wave experiments were conducted in a 6 m long, 0.3 m wide, and 0.6 m high wave flume, with the aim of investigating the breaking processes and the dispersion of oil spills by breaking waves \citep{li2017size,wei2018chaos,afshar2018laboratory}. The breaking waves are initialized by driving a piston-type wavemaker over a constant water depth $d$. A single-wave breaking event is produced by a single push of the wavemaker, and its trajectory $x(t)$ and associated wave plate velocity $u(t)$ are determined by the following functions:
\begin{equation}
  x(t) = \frac{S}{2}(1 - \cos (2 \upi f t)),0\le t\le \frac{1}{2f}
\end{equation}
\begin{equation}
  u(t) = S\upi f \sin (2 \upi f t),0\le t\le \frac{1}{2f}
\end{equation}
where $S$ is the wavemaker stroke length; $f$ is the frequency; and $t$ is the time. A single push of the wavemaker for a half period $1/(2f)$ is applied  to produce a wave with a single crest. During the motion of the wave plate, the maximum wave plate stroke is $S$, and the maximum wave plate velocity is $U=S\upi f$. Multiple types of waves could be generated by varying the stroke $S$, frequency $f$ and water depth $d$, ranging from nonbreaking regular waves to breakers with different intensities. In comparison with the conventional motion of the piston-type wavemaker that produces sinusoidal waves with oscillatory motion of $x(t) = (S/2) \sin(2 \upi f t)$, the piston trajectory here can steepen the wave profile and promote the wave to break. The origin of the experimental domain is located at the undisturbed water surface on the left boundary, where $x$ represents the streamwise direction, and $z$ is the vertical direction, with right and upwards being positive. The wavemaker is initially located at $x=0.535$ m from the left boundary (see figure \ref{fig:domain}). High-speed imaging is implemented for visualizing the wave impingement and the subsequent breakup processes during wave breaking. The turbulence produced by breaking is characterized using particle image velocimetry (PIV). The PIV images are processed to calculate the time evolution of turbulence in the wave tank. Digital inline holography, a 3D imaging technique, is employed to measure the size of the produced droplets and bubbles and to qualify the subsurface particle size distribution. 

On the basis of laboratory experiments, 2D simulations of a range of breaking waves are conducted using the Basilisk solver. Three different breaking waves are simulated to reproduce the breaking characteristics numerically. The wave plate stroke $S$, frequency $f$, and water depth $d$ for generating the three breakers are summarized in table \ref{tab:Sfd}. One of the breakers, a typical plunging breaker with $S=0.5334$ m and $f=0.75$ Hz, is chosen for model verification and detailed analysis. Furthermore, a parametric study is performed to relate the wave characteristics to the initial conditions by extensively varying the stroke $S$, frequency $f$, and water depth $d$.

\begin{table}
  \begin{center}
\def~{\hphantom{0}}
  \begin{tabular}{cccc}
     Wave  &  $S$(m)  &  $f$(Hz)  &  $d$(m) \\ [3pt]
        1    &   0.5334    &   0.75       &   0.25    \\
        2    &   0.4572    &   0.75       &   0.25    \\
        3    &   0.4572    &   0.625      &   0.25    \\
  \end{tabular}
   \caption{Wave plate stroke $S$, frequency $f$, and water depth $d$ for generating three different breaking waves}
  \label{tab:Sfd}
  \end{center}
\end{table}

\subsection{Dimensional analysis for waves generated by a wave plate}\label{sec:Dimensional analysis}
In this section, a dimensional analysis for the waves generated by wave plates is performed. Considering a 2D wave, the wave generated by the wave plate is assumed to be dependent on the fluid properties and the initial conditions. If the wave process is restricted to air-water systems close to standard temperature and pressure, then the density and kinematic viscosity ratios of the two phases are those of air and water in the experiments, which will not be regarded as altering the wave features. Then, the dependent variables for identifying this specific wave can be expressed as follows:
\begin{equation}
f(g, \nu, \rho, \sigma, S, f, d)
  \label{dependent variables}
\end{equation}
where $g$ [dimension $\rm{L/T^2}$] is the gravitational acceleration, $\nu$ [$\rm{L^2/T}$] is the water kinematic viscosity, $\rho$ [$\rm{M/L^3}$] is the water density, and $\sigma$ [$\rm{M/T^2}$] is the surface tension. The piston stroke $S$ [$\rm{L}$] and frequency $f$ [$\rm{T^{-1}}$] of the wave plate, and the undisturbed depth of water $d$ [$\rm{L}$] are referred to as the initial conditions.
Buckingham's theorem can be used to construct the following dimensionless parameters by selecting $\rho$, $g$, and $d$ as the repeating variables:
\begin{equation}
\frac{g^{1/2} d^{3/2}}{\nu}=\Rey, \hspace{3mm} \frac{\rho g d^2}{\sigma}=Bo, \hspace{3mm} \frac{S}{d}, \hspace{3mm} \frac{f}{\sqrt{g/d}}
  \label{dimensionless parameters for wave characteristics}
\end{equation}
The above dimensional analysis indicates that wave characteristics are determined by the Reynolds number $\Rey$, Bond number $Bo$, dimensionless initial stroke $S/d$ and frequency $f / \sqrt{g/d}$ of the wave plate. Of particular interest in this study is the maximum wave height before breaking $H$ [$\rm{L}$], the breaking wave crest $H_b$ [$\rm{L}$] of the plunging breaker, the total energy per unit length transferred by the motion of wave plate $E_l$ [$\rm{ML/T^2}$], and the dissipation of the wave energy per unit length of the breaking crest, $\epsilon_l$ [$\rm{ML/T^3}$]. These wave characteristics should be dimensionless to connect to the dimensionless parameters representing the fluid properties and the initial conditions in (\ref{dimensionless parameters for wave characteristics}). Using dimensional analysis, the dimensionless parameters for these wave features are as follows:
\begin{equation}
\frac{H}{d}, \hspace{3mm} \frac{H_b}{d}, \hspace{3mm} \frac{E_l}{\rho g d^3}, \hspace{3mm} \frac{\epsilon_l}{\rho g^{3/2} d^{5/2}}
  \label{dimensionless parameters of wave characteristics}
\end{equation}
 Quantifying the influence of these dimensionless parameters is of great significance for elucidating the wave shape evolution, energy transfer, and air entrainment mechanisms.
 
\section{Numerical investigation}\label{sec:Numerical study}
\subsection{Basilisk solver}\label{sec:Basilisk solver}
The Navier-Stokes equations for incompressible gas-liquid two-phase flow with variable density and surface tension are simulated using the Basilisk library. The Basilisk package, developed as the successor to the Gerris framework \citep{popinet2003gerris, popinet2009accurate}, is an open-source program for solving various systems of partial differential equations on regular adaptive Cartesian meshes with second-order spatial and temporal accuracy. A quadtree-based adaptive mesh refinement (AMR) scheme is used in 2D calculations to improve computational efficiency by concentrating computational resources on important solution domains. The generic time loop is  implemented in the numerical scheme and the timestep is limited by the Courant–Friedrichs–Lewy (CFL) condition. The incompressible, variable density Navier-Stokes equations with surface tension can be written as:
\begin{equation}
  \rho(\p _t\boldsymbol{u}+(\boldsymbol{u}\bcdot\nabla)\boldsymbol{u})=-\nabla p+\nabla\bcdot(2\mu\mathsfbi{D})+\boldsymbol{f}_\sigma
  \label{N-S1}
\end{equation}
\begin{equation}
  \p _t\rho+\nabla\bcdot(\rho\boldsymbol{u})=0
  \label{N-S2}
\end{equation}
\begin{equation}
  \nabla\bcdot\boldsymbol{u}=0
  \label{N-S3}
\end{equation}
where $\boldsymbol{u}=(u,v,w)$ is the fluid velocity, $\rho\equiv\rho(x,t)$ is the fluid density, $p$ is the pressure, $\mu\equiv\mu(x,t)$ is the dynamic viscosity, $\mathsfbi{D}$ is the deformation tensor defined as $D_{ij}\equiv(\p _i u_j+\p _j u_i)/2$, and $\boldsymbol{f}_\sigma$ is the surface tension force per unit volume \citep{deike2016air}.

The liquid-gas interface is tracked by the momentum-conserving volume-of-fluid (MCVOF) advection scheme \citep{fuster2018all}, while the corresponding volume fraction field is solved by a piecewise linear interface construction (PLIC) approach \citep{scardovelli1999direct,scardovelli2000analytical} with the interface normal being computed by the Mixed-Youngs-Centered (MYC) method \citep{aulisa2007interface}. The volume-of-fluid (VOF) method was originally developed by \citet{hirt1981volume} and has been modified by \citet{kothe1991ripple}, and further coupled with momentum conservation by \citet{fuster2018all}, with the advantage of allowing variable spatial resolution and sharp representation along the interface while restricting the appearance of spurious numerical parasitic currents \citep{zhang2020modeling}. The interface of two-phase flow is reconstructed by a function $\alpha(x,t)$, defined as the volume fraction of a given fluid in each cell of the computational mesh, assuming values of 0 or 1 for each phase. The density and viscosity can thus be computed by arithmetic means as:
\begin{equation}
  \rho(\alpha)=\alpha\rho_1+(1-\alpha)\rho_2
  \label{VOF1}
\end{equation}
\begin{equation}
  \mu(\alpha)=\alpha\mu_1+(1-\alpha)\mu_2
  \label{VOF2}
\end{equation}
where $\rho_1$ and $\rho_2$, $\mu_1$ and $\mu_2$ are the density and viscosity of the first and second fluids, respectively.

An equivalent advection equation for the volume fraction can be obtained by replacing the advection equation for the density:
\begin{equation}
  \p _t \alpha+\nabla\bcdot(\alpha \boldsymbol{u})=0
  \label{advection_VOF}
\end{equation}

A momentum conserving scheme is applied in the advective momentum fluxes near the interface to reduce numerical momentum transfer through the interface. Total fluxes on each face are obtained by adding the diffusive flux due to the viscous term, which are computed by the semi-implicit Crank-Nicholson scheme \citep{pairetti2018bag}. The Bell-Collela-Glaz (BCG) second-order upwind scheme is used for the reconstruction of the liquid and gas momentum per unit volume to be advected in the cell \citep{bell1989second}.

Surface tension is treated with the method of \citet{brackbill1992continuum} and the balanced-force technique \citep{francois2006balanced} as further developed by  \citet{popinet2009accurate, popinet2018numerical}. A generalized version of the height-function (HF) curvature estimation is implemented to address the inconsistency at low interface resolution, giving accurate and efficient solutions for surface-tension-driven flows. The surface tension force per unit volume $\boldsymbol{f}_\sigma$ can be expressed as:
\begin{equation}
  \boldsymbol{f}_\sigma=\sigma\kappa\delta_s\boldsymbol{n}
  \label{surface_tension_force}
\end{equation}
where $\sigma$ is the surface tension coefficient; $\delta_s$ is the interface Dirac function, indicating that the surface tension term is concentrated on the interface; and $\kappa$ and $\boldsymbol{n}$ are the curvature and normal to the interface, respectively.

The integrals over the entire water phase are performed numerically to identify the energy budget in the water. The kinetic energy $E_k$ and the gravitational potential energy $E_p$ of the propagating wave are provided as follows:
\begin{equation}
  E_k=\frac{1}{2}\int_V\rho|\boldsymbol{u}\bcdot\boldsymbol{u}|\mathrm{d} V
  \label{kinetic_energy}
\end{equation}
\begin{equation}
  E_p=\int_V\rho gy\mathrm{d} V - E_{p0}
  \label{gravitional_potential_energy}
\end{equation}
where $V$ is the domain occupied by water in the system and $E_{p0}$ is the gravitational potential energy of the still water at the beginning.
The mechanical energy $E_m$ of the wave is calculated as the sum of the kinetic and potential components:
\begin{equation}
  E_m=E_k+E_p
  \label{mechanical energy}
\end{equation}
The nonconservative energy dissipation from the action of viscosity, $E_d$, can be calculated directly from the deformation tensor:
\begin{equation}
  E_d(t)=\int_0^t\int_v\mu\frac{\p u_i}{\p x_j}\frac{\p u_j}{\p x_i}\mathrm{d} V\mathrm{d}t
  \label{viscous_dissipation}
\end{equation}
Thus, the total conserved energy budget is given by $E_t=E_k+E_p+E_d$.


The bubbles/droplets are identified by tagging connected neighbourhoods. The volume and position of individual bubbles and droplets can be determined by considering connected regions separated by interfacial cells. The advection errors of VOF interface reconstruction can be significant when the liquid structure is less than twice the grid spacing \citep{li2010towards}. As noted in previous studies \citep{wang2016high,mostert2022high}, with more than 4 minimum computational cells per bubble/droplet diameter, the bubbles/droplets take on a spherical shape, consistent with the physical shape of bubbles/droplets with a diameter of less than 1 mm when surface tension dominates \citep{clift2005bubbles}. So in this study,  bubbles/droplets are considered unsolved and will not be counted when their diameters are less than 4 computational cells. 

\subsection{Numerical setup}\label{sec:Numerical setup}
The numerical methodology employed in this investigation involves the simulation of the incompressible flow of two immiscible fluids. To accurately capture the physical features of the wave profiles, the Navier-Stokes equations are solved numerically on sufficiently fine grids so that viscous and capillary effects can be retained. 
Gravity is taken into account using the “reduced gravity approach” \citep{wroniszewski2014benchmarking} by re-expressing gravity in two-phase flows as an interfacial force. An initial depth of water $d$ is used in a square box with a side length of $L=24d=6$ m. The wave propagates in the $x$ direction from left to right. The density and kinematic viscosity ratios of the two phases are those of air and water in the experiments, which are $1.29/1018.3$ and $1.39e-5/1.01e-6$, respectively. The Reynolds number in the breaking wave event generated by the wave plate can be defined by $\Rey=g^{1/2}d^{3/2}/\nu = c_0 d/\nu$, where $c_0 = \sqrt{gd}$ is the linear speed. Due to the limitation of computational resources, combined with the decreasing effects of the Reynolds number on the evolution of wave breaking \citep{mostert2020inertial}, it is possible to use a Reynolds number that is smaller than the actual value. For the plunging breaking wave with $S=0.5334$ m and $f=0.75$ Hz at a water depth of $d=0.25$ m, $\Rey=6 \times 10^4$ is utilized, which corresponds to a water depth of 0.076 m and wave plate stoke of 0.216 m; these values are smaller than the actual values by 3 orders of magnitude. The basic nature of the breaking processes is not expected to be fundamentally altered by Reynolds number effects. The surface tension can be expressed by the Bond number $Bo=\rho g d^2/\sigma$, where $\sigma$ is the constant surface tension coefficient between water and air. The physical value of the water surface tension coefficient with air, $\sigma=0.0728$ $\rm{kg/s^2}$, is used to analyse the effect of surface tension on the formation of the main cavity, resulting in $Bo=8600$.

The numerical resolution is given by $\Delta=L/2^{l_{max}}$, where $l_{max}$ is the maximum level of refinement in the AMR scheme. Three sets of the maximum level of refinement used for mesh convergence analysis in this study are 13, 14, and 15, corresponding to the minimum mesh sizes of 0.732 mm, 0.366 mm, and 0.183 mm, respectively. As the surface tension scheme is time-explicit, the maximum timestep is the oscillation period of the smallest capillary wave. For the maximum level of refinement $l_{max}=15$, the corresponding maximum timestep should not be larger than $6.4e-5$. A CFL number of 0.5 is utilized to ensure numerical stability. VOF tracers are used to capture the water-air interfaces and the moving boundary of the wave plate. This capability of local dynamic grid refinement significantly reduces the computational cost of representing a breaking wave that propagates within an extended physical domain at a high resolution. This makes it especially appropriate for the present application where wave profile evolution and wave breaking are expected. Since the moving piston is updated at each timestep, the grids intersected with the piston are refined to the finest level all the time, thus ensuring the accurate representation of the moving boundary in the adaptive meshes. The refinement criterion is based on the wavelet-estimated discretization error in terms of the velocity and VOF fields. The corresponding mesh will be refined as required when initializing the wave. The wave plate boundary and the air-water interface are initially refined to the finest level, while the remainder of the domain remains at a level of refinement of 10. The refinement algorithm is invoked every timestep to refine the mesh when the wavelet estimated error exceeds $u_{err} = 1e-2$ for the velocity field and $f_{err} = 1e-6$ for the volume fraction field.

Breaking waves are normalized using the reference length and velocity scales, which in this case are the still water depth $d$ and wave celerity $c=\sqrt{g(H+d)}$, respectively; the reference time scale is defined as $t_0 = d/c_0 = \sqrt{d/g}$.
\subsection{Mesh convergence}\label{sec:Mesh convergence}
\begin{figure}
  \centering
  \begin{overpic}[width=\linewidth]{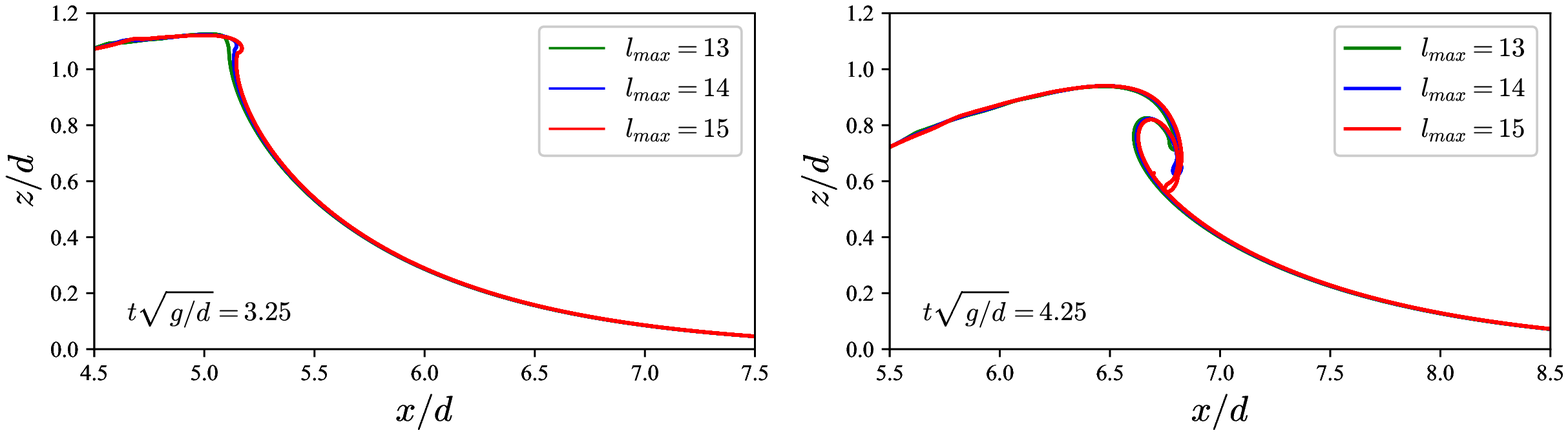}
  \put(0,26){\small(\textit{a})}
  \put(50,26){\small(\textit{b})}
  \end{overpic}
  ~
  \begin{overpic}[width=\linewidth]{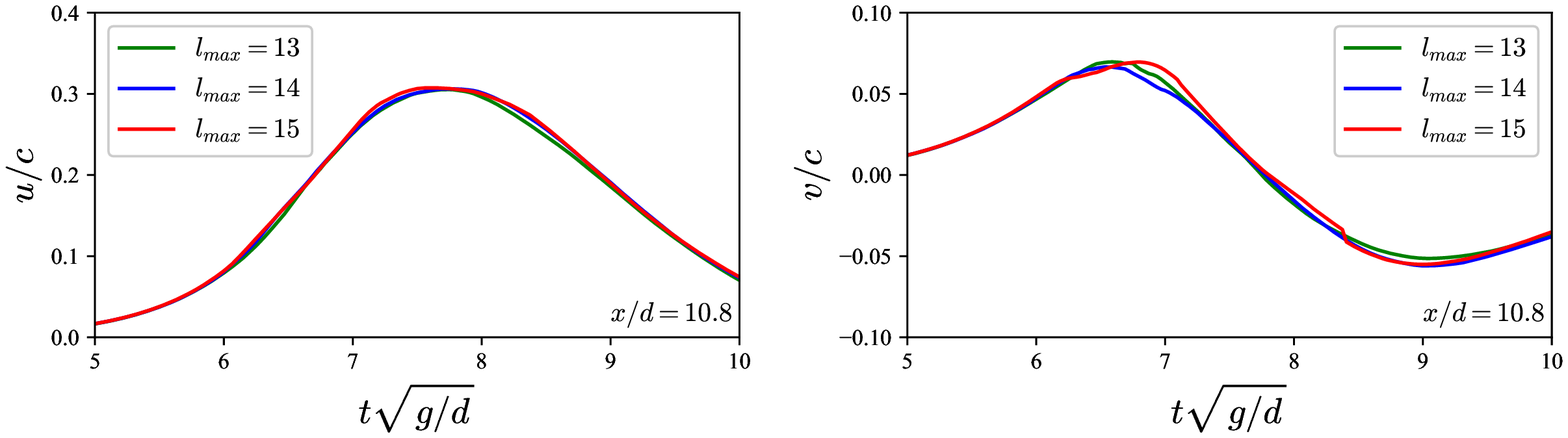}
  \put(0,26){\small(\textit{c})}
  \put(50,26){\small(\textit{d})}
  \end{overpic}
  ~
  \begin{overpic}[width=\linewidth]{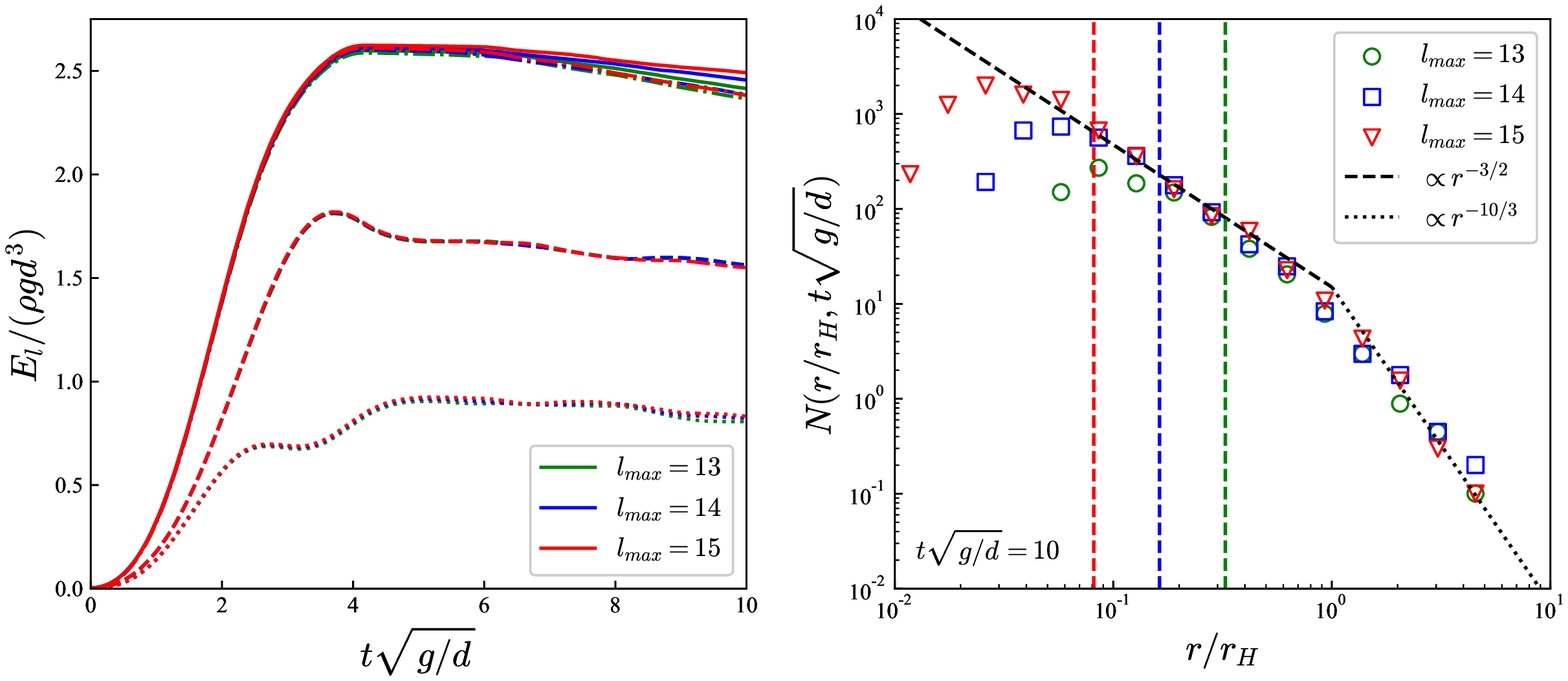}
  \put(0,41){\small(\textit{e})}
  \put(50,41){\small(\textit{f})}
  \end{overpic}
  \caption{Convergence study at three different mesh resolutions for wave 1 with $S/d$ = 2.13, $f / \sqrt{g/d}$ = 0.12, green, $2^{13}$; blue: $2^{14}$; red: $2^{15}$. Grid convergence of free surface during wave breaking at $t \sqrt{g/d}$ = 3.25 (a), and jet impinging at $t \sqrt{g/d}$ = 4.25 (b); the temporal evolution for horizontal component (c) and vertical component (d) of velocity field in the broken bore propagation region at $x/d$ = 10.8; the energy budget (e) for kinetic energy $E_k$ (dotted), the gravitational potential energy $E_p$ (dashed), the mechanical energy $E_m$ (dashdot), and the total conserved energy $E_t$ (solid); and bubble size distribution (f) in the late wave breaking stage at $t \sqrt{g/d}$ = 10, the black dashed line represents the scaling law of -3/2 for $r < r_H$ and the black dotted line represents the scaling law of -10/3 for $r > r_H$. The red, blue, and green dashed lines represent the diameter of bubbles with 4 computational cells, and to the left of these lines, bubbles are considered unsolved.}
\label{fig:GridConvergence}
\end{figure}
The choice of the effective numerical resolution is related to the numerical convergence. A key physical feature of simulating two-phase breaking waves is the thickness $\delta$ of the viscous boundary layer at the free surface. The estimation from Batchelor’s method suggests the defining length scale $\delta \sim d/\sqrt{\Rey} \approx 0.004d = 1.0$ mm (Deike et al. 2015, 2016). Based on this estimation, the viscous sublayer is resolved with more than five grid cells at $l_{max}$ = 15, allowing us to resolve the dissipation rate associated with the breaking waves \citep{mostert2022high}.
Furthermore, the grid convergence of the numerical results is analysed by considering three sets of simulations with $l_{max}$ = 13, 14, and 15, corresponding to the effective resolution, which is equivalent to $4096^2$, $8192^2$ and $16384^2$ on a regular grid, respectively. The numerical convergence is discussed in terms of the evolution of the free surface, velocity field, energy budget, and size distribution of the bubbles entrapped by wave breaking. Figure \ref{fig:GridConvergence}(a,b) shows the influence of the mesh resolution on the free-surface development of wave 1. The wave breaks at $t \sqrt{g/d}$ = 3.25, at which the front of the wave has a vertical slope (a), and increasingly smaller differences can be observed at the tip of the overturning wave with $l_{max}$ increasing from 13 to 15. The overturning jet curls over itself and impacts the surface of the wave front at $t \sqrt{g/d}$ = 4.25 (b). A slight phase shift can be seen at different resolutions, but the entrained air by the plunging jet is quite similar in both shape and size. Next, figure \ref{fig:GridConvergence}(c,d) shows the temporal evolution of the horizontal component (c) and vertical component (d) of the velocity field in the broken-bore propagation region at $x/d$ = 10.8. This demonstrates a better agreement between the cases with resolutions of $2^{14}$ and $2^{15}$ compared to those between $2^{13}$ and $2^{14}$. For all cases in figure \ref{fig:GridConvergence}(e), the results of the energy budget converge for the evolution of kinetic energy $E_k$, gravitational potential energy $E_p$, and conservative energy $E_m = E_k + E_p$, indicating that numerical convergence is achieved during the energy transfer between $E_k$ and $E_p$. The differences in $E_t = E_k + E_p + E_d$ at different resolutions indicate that grid cells cannot fully capture the dissipated energy directly; however, since the wave dissipation rate can be calculated based on the conservative energy $E_m$, numerical convergence is also achieved regarding the energy dissipation calculated as the loss of $E_m$. Another indicator of numerical convergence for breaking waves is bubble generation due to air entrainment by breaking. Figure \ref{fig:GridConvergence}(f) shows the bubble size distribution in the late stage after wave breaking at $t \sqrt{g/d}$ = 10. The smaller-sized bubbles can be resolved and captured when increasing the mesh resolution. All cases collapse onto a similar curve, which follows the scaling law of -3/2 for $r < r_H$ and -10/3 for $r > r_H$, where $r_H$ is the critical Hinze scale, which is approximately 4.5mm at this time for wave 1 \citep{hinze1955fundamentals,deane2002scale,mostert2022high}.

The above convergence studies confirmed that all results are well converged and no significant changes are observed when the maximum level of refinement increases from 13 to 15. The resolution of $2^{15}$ is used to realize a more precise parametric study to determine the wave characteristics as a function of the fluid properties and initial conditions; thus, all results presented below have converged regarding the grid resolution.

\subsection{Breaking wave verification}\label{sec:model verification}
\begin{figure}
  \centering
  \begin{overpic}[width=\linewidth]{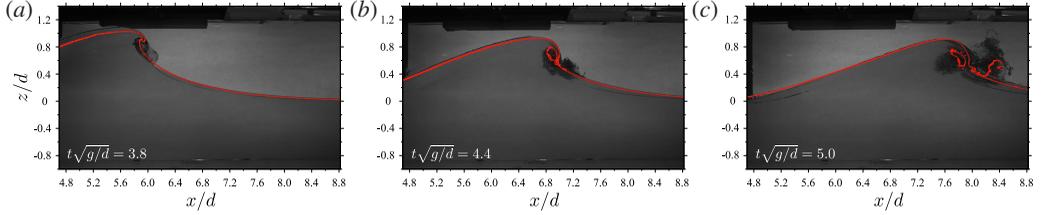} 
  \put(-0.75,19){\small(\textit{a})}
  \put(33,19){\small(\textit{b})}
  \put(66.5,19){\small(\textit{c})}
  \end{overpic}
  \caption{Qualitative comparison of free surface profiles between laboratory images and numerical results for wave 1 with $S/d$ = 2.13, $f / \sqrt{g/d}$ = 0.12.}
\label{fig:profiles}
\end{figure}
A high-speed camera with a frame rate of 500 frames per second is used in the experiments to visualize the development of wave breaking and subsequent breakup processes. The field of view, 103 $\times$ 103, is centred horizontally at $x/d=6.64$. The vertical centre of the camera is adjusted to the initial free surface. The numerical results of the temporal evolution of the free surface for wave $1$ are compared with experimental snapshots for model verification. Comparisons of the free-surface profile between the simulation results and snapshots taken during the experiments are shown in figure \ref{fig:profiles}.
The camera is located upstream of the wave direction close to the side of the wave plate. This device is primarily responsible for recording the development of the plunging jet, jet impact and air entrainment, and the generation of the first splash-up. Comparisons of the free-surface evolution at $t \sqrt{g/d}$ = 3.8, 4.4, and 5.0 show excellent agreement between the current simulation and the experimental results. As the wave slope becomes steeper and the wave crest curls over, the plunging jet can be observed at $t \sqrt{g/d}$ = 3.8, with the tendency to project downwards to the water surface. At $t \sqrt{g/d}$ = 4.4, the plunging jet impacts the rising wave front, forming the main cavity by entrapping a tube of air. Driven by the primary plunging jet, a splash-up is produced moving upwards at $t \sqrt{g/d}$ = 5.0, and some droplets can be observed from fractured ligaments. A small discrepancy between the height of the splash-up and the development of the aerated region is explained by the air entrainment caused by the 3D instability in the spanwise direction, which is outside the purview of the present study. Overall, the evolution of the free surface during this process, including the curvature of the overturning wave crest, the size of the main cavity, and the height and location of the first splash-up, can be accurately predicted by our numerical simulations.

\begin{figure}
  \centering
  \begin{overpic}[width=0.8\linewidth]{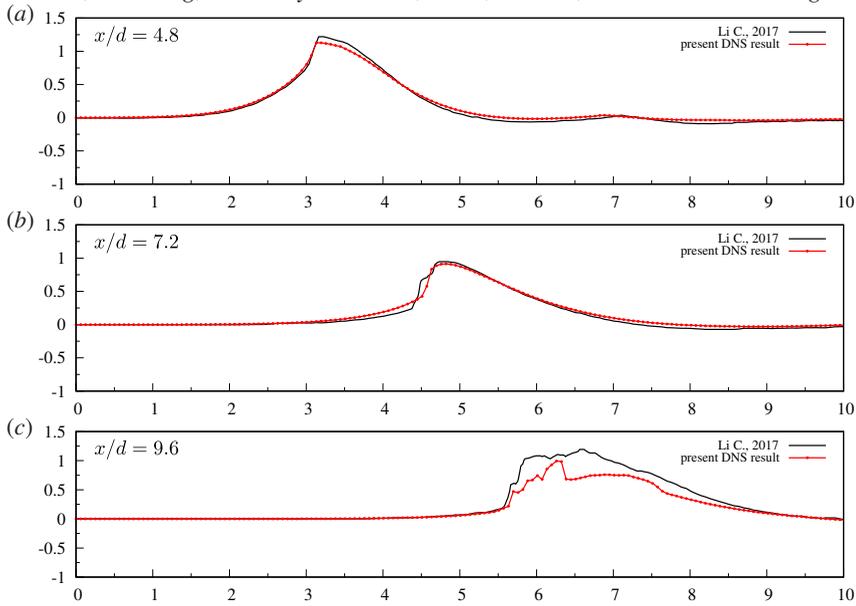} 
  \put(-4,71){\small(\textit{a})}
  \put(-4,45.75){\small(\textit{b})}
  \put(-4,20.5){\small(\textit{c})}
  \end{overpic}
  \caption{Qualitative comparison of surface elevations over time at $x/d$ = 4.8 (a), 7.2 (b), and 9.6 (c) for wave 1 with $S/d$ = 2.13, $f / \sqrt{g/d}$ = 0.12.}
\label{fig:interface}
\end{figure}

Moreover, figure \ref{fig:interface} shows the simulated free-surface profiles over time for wave 1 recorded at three designated positions ($x/d$ = 4.8, 7.2, and 9.6) corresponding to the prebreaking, breaking, and postbreaking regions, respectively, with a comparison to the experimental high-speed imaging results.
The free-surface profile at the first position ($x/d$ = 4.8) remains smoothly curved, which corresponds to the prebreaking stage where the free surface is smooth, without the formation of the vertical interface and the generation of bubbles and droplets. The numerical simulation accurately reproduces the evolution of the free surface, including the development of the rise and fall of the wave profile, with only a slight underestimation at the peak value of the wave profile at $t \sqrt{g/d}$ = 3.1.
The second position is located at $x/d$ = 7.2, within the wave-breaking region, near the main cavity entrapped by the plunging jet. In the experiment, the free surface exhibits an immediate increase after jet impact at approximately $t \sqrt{g/d}$ = 4.4, indicating the penetration of the plunging jet into the wave front and the formation of the main cavity. Figure \ref{fig:interface}(b) shows that our numerical simulation can closely capture the phenomenon of how waves break. The only discrepancy can be caused by the lack of small ejections when the plunging jet penetrates into the wave front due to the absence of the 3D effect.
The wave propagates to the third position and develops into turbulent flow, forming a large amount of spray and bubbles. There are apparent fluctuations in the free surface between $t \sqrt{g/d}$ = 5.6 and 8.8, showing a strongly turbulent phenomenon in this region. Figure \ref{fig:interface}(c) shows an overall underestimation of the free-surface elevations from $t \sqrt{g/d}$ = 5.6 to 8.8 by our numerical simulation. This result is most likely due to differences in the recordings of the free-surface elevations from the experiments and numerical simulations. In the experiment, the value of the free-surface elevations is the maximum elevation of the wave profile, splashing bubbles and droplets, as the free-surface elevations are recorded from the black region in the experimental snapshots. However, in the numerical simulation, the free-surface elevations are primarily determined by wave profiles rather than splashing droplets scattered above the water surface.
In general, the temporal evolution of free-surface profiles can be precisely reproduced by our simulation when compared to laboratory experiments at each location.

In summary, despite the limitations of the 2D simulation in producing droplets and ligaments in the spanwise direction, the ability of our model to capture wave hydrodynamics, including accurate reproduction of the wave height, wave speed, and wave-breaking process, can be demonstrated through the comparisons above.


\section{Breaking characteristics}\label{sec:Breaking characteristics}
\subsection{Wave-breaking dynamics}\label{sec:Wave breaking dynamics}
\begin{figure}
  \centering
  \begin{overpic}[width=\linewidth]{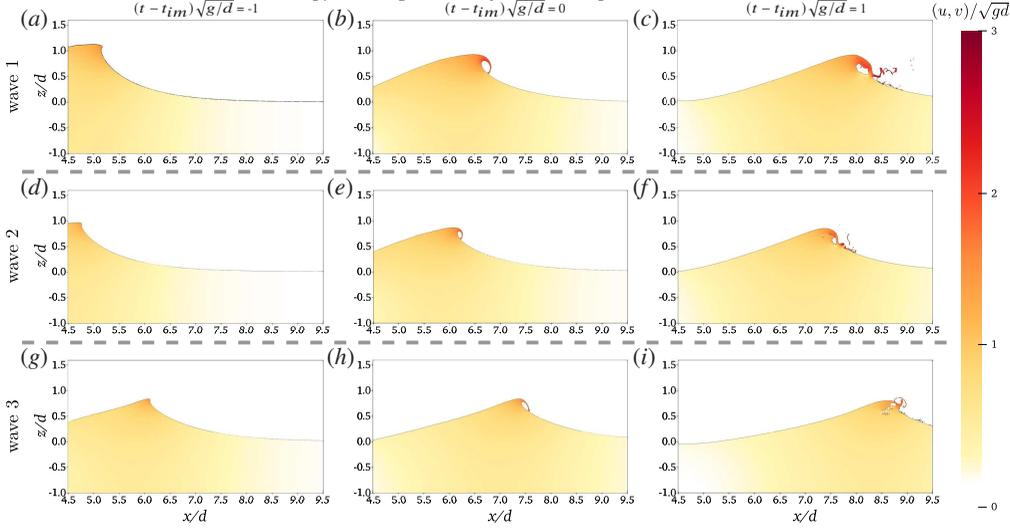} 
  \put(1.5,48.5){\small(\textit{a})}
  \put(1.5,32){\small(\textit{d})}
  \put(1.5,15.5){\small(\textit{g})}
  \put(31.5,48.5){\small(\textit{b})}
  \put(31.5,32){\small(\textit{e})}
  \put(31.5,15.5){\small(\textit{h})}
  \put(61.5,48.5){\small(\textit{c})}
  \put(61.5,32){\small(\textit{f})}
  \put(61.5,15.5){\small(\textit{i})}
  \put(12.5,49.75){\tiny{$(t-{t_{im}}) \sqrt{g/d}$ = -1}}
  \put(43,49.75){\tiny{$(t-{t_{im}}) \sqrt{g/d}$ = 0}}
  \put(72.5,49.75){\tiny{$(t-{t_{im}}) \sqrt{g/d}$ = 1}}
  \end{overpic}
  \caption{Evolution of free surface for the three plunging breakers with different breaking intensity, coloured by the normalized velocity magnitude $(u,v)/c_0$, where $c_0 = \sqrt{gd}$ is the wave phase speed. $t_{im}$ is the time when the plunging jet impacts on the wave front.}
\label{fig:wave123_colormap}
\end{figure}
Sequences of three different plunging breakers with the contours of the normalized velocity magnitude $(u,v)/c_0$ are shown in figure \ref{fig:wave123_colormap}, where $c_0 = \sqrt{gd}$ is the linear speed. The parameter $t_{im}$ represents the time when the plunging jet impacts the wave front. For wave 1, the wave begins to break as the wave crest steepens and becomes multivalued at $(t-{t_{im}}_1) \sqrt{g/d} = -1$. A curled jet is formed projecting ahead of the wave, and a high and flat interface accumulates at the backside of the wave crest. The overturning jet develops further and impacts the wave front, forming a closed cavity from the entrapped air at $(t-{t_{im}}_1) \sqrt{g/d} = 0$, with the production of a splash-up at $(t-{t_{im}}_1) \sqrt{g/d} = 1$. The phenomena of the breaking event from wave breaking and jet impacting to splash-up formation among waves 1, 2, and 3 are quite similar. However, some differences exist at the backside of the wave crest and regarding the size and shape of the closed cavity.
\begin{figure}
  \centering
  \begin{overpic}[width=\linewidth]{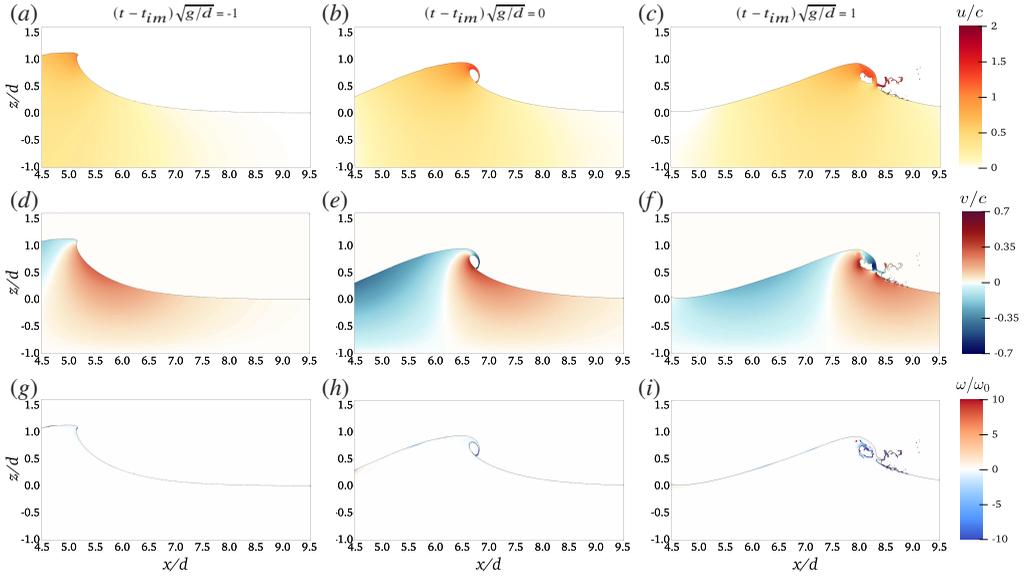}
  \put(0.5,53.75){\small(\textit{a})}
  \put(31,53.75){\small(\textit{b})}
  \put(62,53.75){\small(\textit{c})}
  \put(0.5,35.5){\small(\textit{d})}
  \put(31,35.5){\small(\textit{e})}
  \put(62,35.5){\small(\textit{f})}
  \put(0.5,17.25){\small(\textit{g})}
  \put(31,17.25){\small(\textit{h})}
  \put(62,17.25){\small(\textit{i})}
  \put(10.5,54){\tiny{$(t-{t_{im}}) \sqrt{g/d}$ = -1}}
  \put(41,54){\tiny{$(t-{t_{im}}) \sqrt{g/d}$ = 0}}
  \put(71.5,54){\tiny{$(t-{t_{im}}) \sqrt{g/d}$ = 1}}
  \end{overpic}
  \caption{Detailed normalized streamwise velocity $u/c$ (a-c), vertical velocity $v/c$ (d-f), and vorticity $\omega / \omega_0$ (g-i) during wave overturning (left column, $(t-{t_{im}}) \sqrt{g/d}$ = -1), jet impinging (middle column, $(t-{t_{im}}) \sqrt{g/d}$ = 0), and splash-up (right column, $(t-{t_{im}}) \sqrt{g/d}$ = 1).}
\label{fig:wave1_colormap_a}
\end{figure}

The details of the break process are demonstrated using wave 1 as an example. Figure \ref{fig:wave1_colormap_a} shows the normalized streamwise velocity $u/c$, vertical velocity $v/c$, and vorticity $\omega / \omega_0$ during wave overturning (left column, $(t-{t_{im}}) \sqrt{g/d}$ = -1), jet impingement (middle column, $(t-{t_{im}}) \sqrt{g/d}$ = 0), and splash-up (right column, $(t-{t_{im}}) \sqrt{g/d}$ = 1), where $c=\sqrt{g(H+d)}$ and $\omega_0=\sqrt{g/(H+d)}$. Figure \ref{fig:wave1_colormap_a}(a-c) shows that the streamwise velocity component of the overturning jet is the largest and begins to accelerate after wave breaking, with the maximum $u/c$ = 1.1 (a), 1.5 (b), and 1.8 (c). Combined with the distribution of the vertical velocity, the water-particle velocities of the wave crest are found to be approximately vertical, as shown by PIV measurements of breaking waves by \citep{perlin1996experimental}. The vertical asymmetry can be clearly observed from the distribution of the vertical velocity. Vortices are identified as concentrated at the free surface as the wave overturns, becoming more intense during cavity closure and subsequent splash-ups.
\begin{figure}
  \centering
  \begin{overpic}[width=\linewidth]{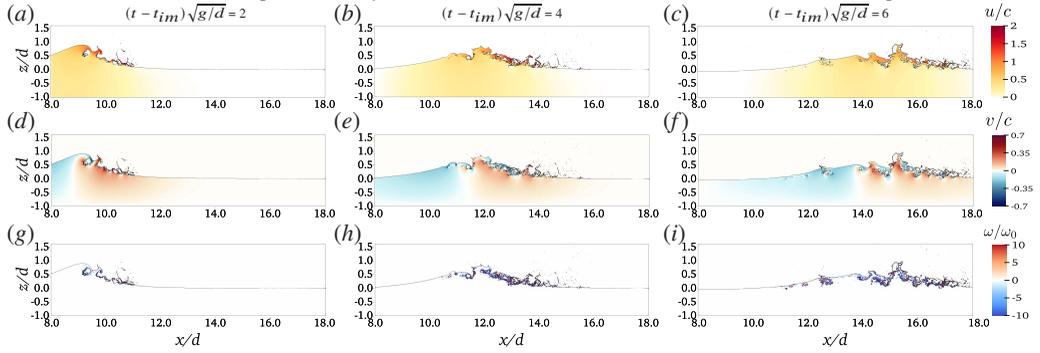}
  \put(-0.5,32){\small(\textit{a})}
  \put(31.5,32){\small(\textit{b})}
  \put(63.75,32){\small(\textit{c})}
  \put(-0.5,21.5){\small(\textit{d})}
  \put(31.5,21.5){\small(\textit{e})}
  \put(63.75,21.5){\small(\textit{f})}
  \put(-0.5,10.5){\small(\textit{g})}
  \put(31.5,10.5){\small(\textit{h})}
  \put(63.75,10.5){\small(\textit{i})}
  \put(11,32.25){\tiny{$(t-{t_{im}}) \sqrt{g/d}$ = 2}}
  \put(42,32.25){\tiny{$(t-{t_{im}}) \sqrt{g/d}$ = 4}}
  \put(74,32.25){\tiny{$(t-{t_{im}}) \sqrt{g/d}$ = 6}}
  \end{overpic}
  \caption{Detailed normalized streamwise velocity $u/c$, vertical velocity $v/c$, and vorticity $\omega / \omega_0$ in the late stage after wave breaking at $(t-{t_{im}}) \sqrt{g/d}$ = 2, 4 and 6.}
\label{fig:wave1_colormap_b}
\end{figure}
Figure \ref{fig:wave1_colormap_b} shows the normalized streamwise velocity $u/c$, vertical velocity $v/c$, and vorticity $\omega / \omega_0$ in the late stage after wave breaking at $(t-{t_{im}}) \sqrt{g/d}$ = 2, 4 and 6. The largest streamwise velocity components concentrate on the ruptured ligaments and ejected droplets from the splash-ups, with a maximum value of $u/c$ = 1.8, as shown in figure \ref{fig:wave1_colormap_b}(a-c). The location of the original wave crest and the number and location of the main splash-up processes can be identified from the distribution of the vertical velocity component in figure \ref{fig:wave1_colormap_b}(d-f) since $v/c$ is equal to 0 at the position of the original wave crest and impacting point. After repetitive and numerous processes of jet impact and splash-up, the interfaces in the wave front become turbulent, forming some irregular turbulent patches. Figure \ref{fig:wave1_colormap_b}(g-i) shows that the vortices do not interact with the bottom, indicating that turbulent clouds induced by wave breaking are not impacted by the wave depth for our breaking wave.

\subsection{Energy budget}\label{sec:Energy budget}
\begin{figure}
  \centering
  \begin{overpic}[width=\linewidth]{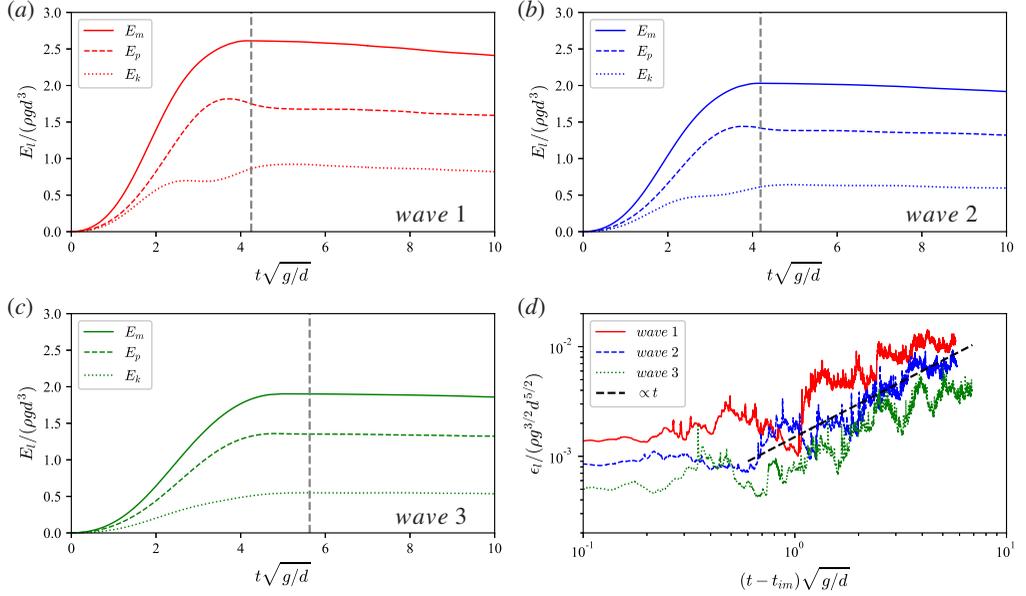}
  \put(0, 57){\small(\textit{a})}
  \put(38, 37){\small{$wave$ $1$}}
  \put(50, 57){\small(\textit{b})}
  \put(88, 37){\small{$wave$ $2$}}
  \put(0, 28){\small(\textit{c})}
  \put(38, 7.5){\small{$wave$ $3$}}
  \put(50, 28){\small(\textit{d})}
  \end{overpic}
  \caption{Temporal evolution of the normalized energy per unit length $E_l/(\rho g d^3)$ for wave 1 (a), wave 2 (b), and wave 3 (c), and the normalized energy dissipation rate per unit length $\epsilon_l/(\rho g^{3/2} d^{5/2})$ (d). The energy has been transferred to the still water column by the motion of wave plate, generating waves propagate in the constant depth of water. Wave breaking enhances the energy dissipation especially near the interface. The jet impacting on the wave front occurs at $t \sqrt{g/d}$ = 4.25, 4.19, and 5.63 for wave 1, 2, and 3, respectively. The grey dashed line represents the moment when jet impacts.}
\label{fig:wave123_E+e}
\end{figure}

Figure \ref{fig:wave123_E+e} presents the time evolution of the components of the energy budget for three cases and a comparison of the energy dissipation. For each case, there is no energy in the system of the wave tank at the beginning; as the wave plate starts to move and act on the water body, the wave can be generated along with increasing gravitational potential energy and kinetic energy. The wave energy continues to increase until the jet strikes the wave front, which occurs at $t \sqrt{g/d}$ = 4.25, 4.19, and 5.63 for waves 1, 2, and 3, respectively. Afterwards, the interface at the wave front can be highly perturbed by the impacting and splashing up of the jet, accompanied by air entrainment and the generation of entrapped bubbles and ejected droplets. These breaking processes enhance the energy transfer and dissipation, resulting in the loss of wave energy. As seen in figure \ref{fig:wave123_E+e}(a) from the energy evolution of wave 1, starting from the initial impact of the plunging jet, there are some visible energy transfers between kinetic and potential energy caused by splash-up production. As the wave breaking progresses, the wave crest diminishes, the plunging jet strikes the free surface and penetrates into the water, $E_{\rm{k}}$ rapidly increases and $E_{\rm{p}}$ begins to decline until the first splash-up occurs at approximately $t \sqrt{g/d}$ = 4.9. The total mechanical energy decays gradually with a continuously increasing decay rate over this breaking phase. In the later stage of breaking waves, notably after $t \sqrt{g/d}$ = 8.25, the wave becomes more turbulent, and the total mechanical energy exhibits a greater decay due to substantial air-water mixing and vortical structures. A similar evolution can be observed in waves 2 and 3, but with a gradual weakening of the energy transfer and turbulence region due to breaking.
Figure \ref{fig:wave123_E+e}(d) exhibits viscous dissipation due to breaking after jet impact. The spanwise width of the numerical domain is regarded as a unit in 2D simulations. As seen from figure \ref{fig:wave123_E+e}(d), the dissipation remains at a roughly constant rate from the moment of impact until $(t-t_{im}) \sqrt{g/d}$= 0.2 or 0.3 after impact, when it begins to increase. Then, the dissipation rate is markedly intermittent, and the occurrences of maximum dissipation rate fluctuations are closely related to the exchange time of energy transfer, i.e. the moments when $E_{\rm{k}}$ and $E_{\rm{p}}$ reach their extreme values. Starting from $(t-t_{im}) \sqrt{g/d}$= 1, the energy dissipation rate appears to follow a linear trend by measuring the energy dissipation of three waves from the moment of breaking on the log-log scale, and then remains constant.
 
\subsection{Air entrainment}\label{sec:Air entrainment}
\begin{figure}
  \centering
  \begin{overpic}[width=\linewidth]{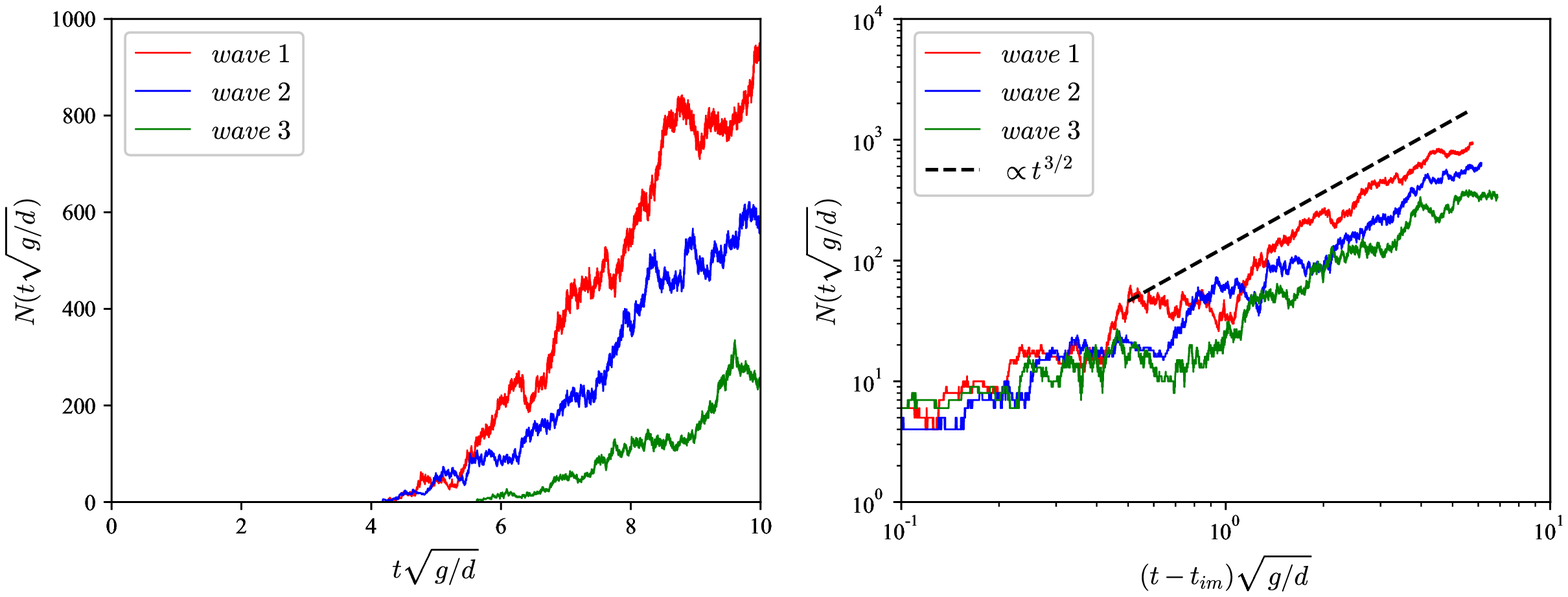}
  \put(0,36){\small(\textit{a})}
  \put(50,36){\small(\textit{b})}
  \end{overpic}
  ~
  \begin{overpic}[width=\linewidth]{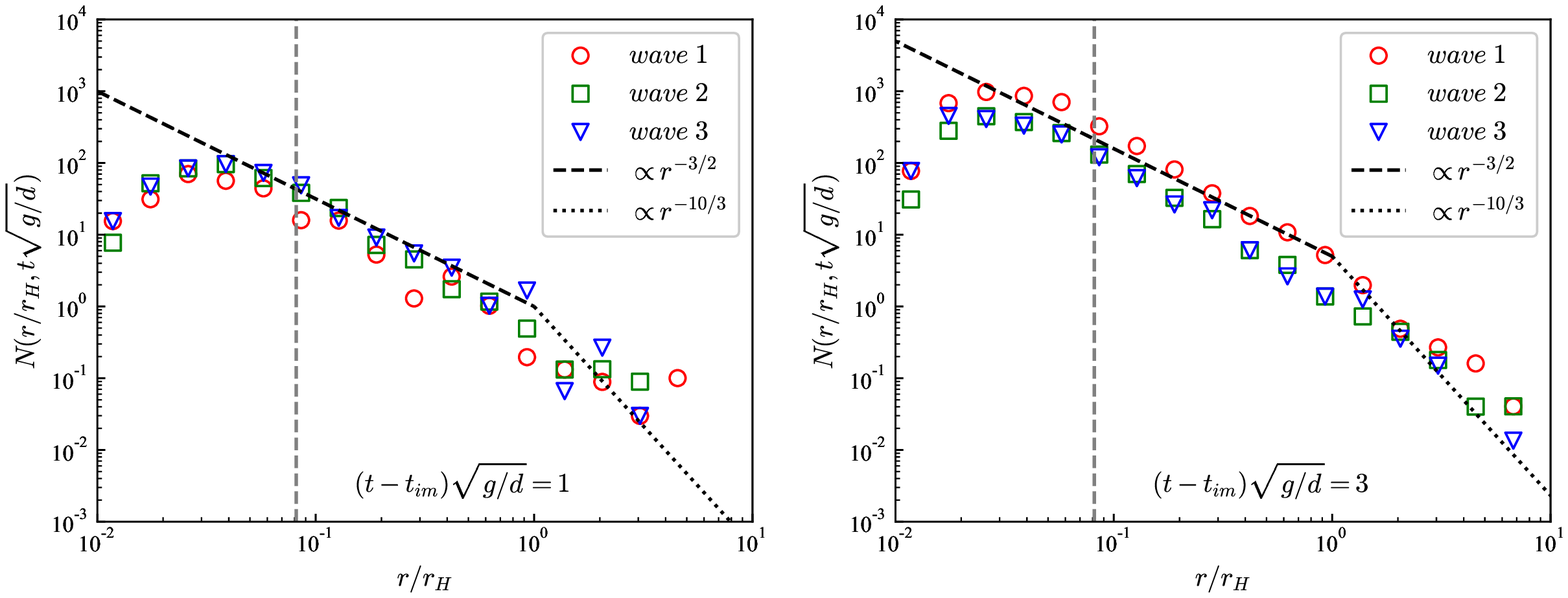}
  \put(0,36){\small(\textit{c})}
  \put(50,36){\small(\textit{d})}
  \end{overpic}
  \caption{Time histories of the number of bubbles (a) and the bubbles number from the moment of breaking on the log-log scale (b) of three different waves. Bubble size distributions at early ($(t-t_{im}) \sqrt{g/d}$ = 1) (c), and late ($(t-t_{im}) \sqrt{g/d}$ = 3) (d) times. $t_{im}$ represents the time when the overturning jet hits the wave front, which corresponds to different times for different waves. Good collapses of the bubble size distributions are shown on the scaling law of -10/3 for super-Hinze regime (dotted), and -3/2 for sub-Hinze regime (dashed), where Hinze scale $r_H$ is roughly 4.5 mm in these cases. The grey dashed lines represent the diameter of bubbles with 4 computational cells, and to the left of these lines, bubbles are considered as unsolved.}
\label{fig:wave123_bubble}
\end{figure}

During wave breaking processes, a large amount of air is injected into the water through jet impingement and splash-up formation, which is distinguished by a wide distribution of bubble sizes. The 2D numerical studies in the breaking-wave literature may not enable us to report an accurate number of bubbles, but the evolution of the bubble formation and breakup processes and the scaling law of the bubble size distribution can be generally captured by fine grid scales through DNS, as shown in figure \ref{fig:wave123_bubble}.

As shown in figure \ref{fig:wave123_bubble}(a), the first bubble is identified at the moment the plunging jet impacts ($t_{im}$), also referred to as the main cavity initially ingested in the breaking process. Subsequently, the first splash-up develops and penetrates into the water, with the main cavity being squeezed and distorted, generating many small bubbles. During this period, the total number of bubbles $N(t \sqrt{g/d})$ begins to increase, but there is no significant increase in the total area of ingested bubbles. There are two sudden increases in the total number of ingested bubbles, which are related to the first and second splash-ups. Similar phenomena can be observed in the three cases. By measuring the bubble number of three waves from the moment of breaking on a log-log scale, the total count as a function of time exhibits a power-law scaling of 3/2, as shown in figure \ref{fig:wave123_bubble}(b). Notably, the temporal evolution of the number of bubbles shows a high similarity to the viscous dissipation rate during the breaking process (see also figure \ref{fig:wave123_E+e}(d)). This implies that there could be a link between the number of bubbles and the energy dissipation rate.

For each case, the number and size of bubbles are sampled at various times and binned by equivalent bubble radius $r$ into bins of size $\Delta r$, resulting in a time-dependent size distribution $N(r/r_H, t \sqrt{g/d})$, where $r_H$ is the Hinze scale, which has been normalized by bin size such that $\int N(r/r_H, t \sqrt{g/d}) dr = \sum N(r/r_H, t \sqrt{g/d}) \Delta r = N(t \sqrt{g/d})$, where $N(t \sqrt{g/d})$ is the total number of bubbles at time $t$ and summation is performed across all radius bins. The bubble size distributions at early ($(t-t_{im}) \sqrt{g/d}$ = 1) and late ($(t-t_{im}) \sqrt{g/d}$ = 3) times are explored in figure \ref{fig:wave123_bubble}(c,d). At both early and late times, the bubble size distributions follow the scaling law of -10/3 for super-Hinze scales and -3/2 for sub-Hinze scales, which is in agreement with previous studies.

\section{Parametric study as a function of the fluid properties and initial conditions}\label{sec:Parametric study}
\subsection{Influence of the Bond number on main cavity}\label{sec:fluid proprieties}
In this section, the effect of dimensionless parameters responsible for the wave evolution and breaking characteristics on the geometry of the main cavity at impact is investigated.
\citet{mostert2022high} indicated that the effect of the Reynolds number on the wave evolution is expected  to be small before wave breaking, as the jet thickness is independent of the Reynolds number, and no apparent dependence of the cavity size on the Reynolds number is discovered. The Reynolds independence of the wave characteristics and main cavity features is checked by comparing the numerical results for distinct Reynolds numbers of $6 \times 10^4$ and $6 \times 10^5$ with experimental data. These finding confirm the results obtained previously by \citet{mostert2022high}. The influence of the Reynolds number on the wave features is neglected in this study since it has been shown to be negligible at high Reynolds numbers in breaking waves.
\begin{figure}
  \centerline{\includegraphics[width=\linewidth]{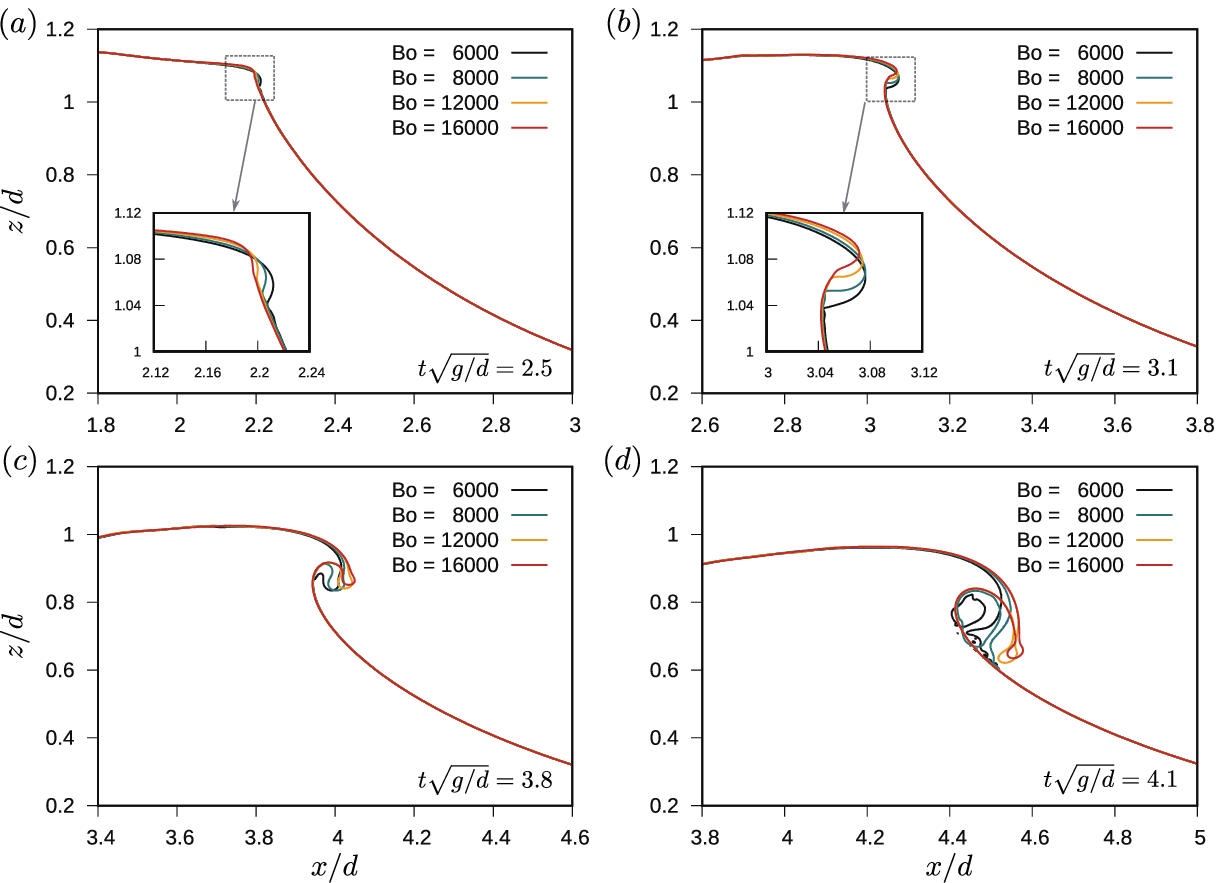}}
  \caption{The spatial evolution of free surface and the development of overturning jet for wave 1 at various Bond numbers when $t \sqrt{g/d}$ = (a) 2.5, (b) 3.1, (c) 3.8, (d) 4.1.}
\label{fig:Bo}
\end{figure}
Since our 2D simulation provides a reasonable estimate of the wave profile and the formation of a plunging jet, which is considered the laminar structure before jet impingement occurs, the effects of the Bond number on the evolution of the wave profile and breaking characteristics of plunging breakers are determined by examining extensive cases with a wide range of Bond numbers. The Bond number increases from $6000$ to $80000$ in increments of $2000$, while all other parameters are constant. Note that $Bo=8.6\times10^3$ refers to the surface tension between air and water in the experiments. Previous studies have revealed that a larger value of $Bo$ results in a greater separation between the wavelength and Hinze scale, necessitating the use of costly numerical resources if all scales are to be resolved \citep{wang2016high}. Our high-resolution meshes that benefit from adaptive mesh refinement criteria can resolve breakers with greater separation between length scales, allowing us to vary $Bo$ over a wide range.

Figure \ref{fig:Bo} shows the evolution of the wave profile under multiple Bond numbers at $t \sqrt{g/d}$ = 2.5, 3.1, 3.8, and 4.1. Qualitatively, there is no significant influence of $Bo$ on the wave profile evolution. The influence of $Bo$ is primarily concentrated on the development of the plunging jet. At $t \sqrt{g/d}$ = 2.5, the generated wave crest is affected by surface tension, producing a bulge at the front face of the steepening wave crest. From the inset of figure \ref{fig:Bo}(a), a smaller bulge can be observed as the Bond number increases, and the wave height before breaking becomes slightly larger. This shows that the effect of surface tension tends to produce capillary ripples at the forward face of the wave, causing a bulge on the water surface. For a larger Bond number at which the effect of surface tension is negligible, a smaller bulge is produced. At $t \sqrt{g/d}$ = 3.1 (figure \ref{fig:Bo}(b)), as the horizontal asymmetry of the wave profile develops further, the edge of the bulge erupts from a point just forward of the crest and becomes tangent to the wave direction, presenting different widths of the bulge due to different surface tensions. The bulge due to surface tension projects forward and develops to the plunging jet at $t \sqrt{g/d}$ = 3.8 (figure \ref{fig:Bo}(c)), and a thicker jet can be observed at a smaller Bond number, indicating that jet thickness is dependent on the Bond number due to capillary effects caused by surface tension. Figure \ref{fig:Bo}(d) shows that at $t \sqrt{g/d}$ = 4.1, the plunging jets at $Bo$ = 6000 and 8000 impact the rising wave front, ingesting a tube of air, while the plunging jets at $Bo$ = 12000 and 16000 still need more time to form the cavity. This is because the thicker jet with heavier mass descends faster under gravity. As the Bond number increases, the instant at which the plunging jet impinges on the front of the wave is delayed, and the plunging jet becomes thinner and projects further forward ahead of the wave, entrapping more air into the wave. 
\begin{figure}
  \centerline{\includegraphics[width=0.75\linewidth]{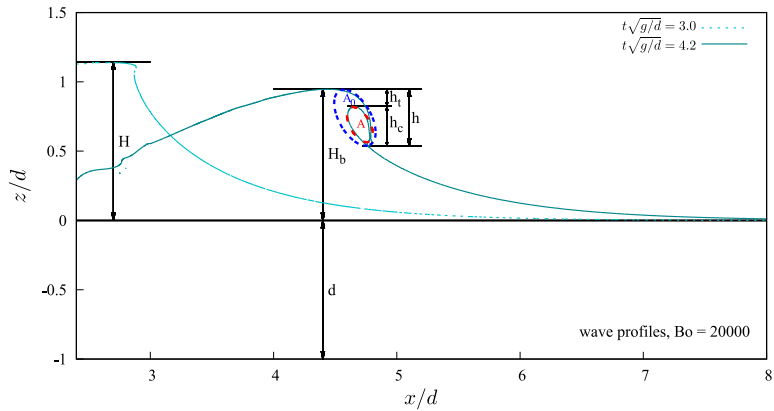}}
  \caption{Estimation of the breaking height $h$, which is the sum of the height from breaking crest to cavity top $h_t$ and the vertical height of the main cavity $h_c$. The main cavity size $A$ is assumed to be proportional to ${h_c}^2$, which can be normalized by $A_0 \propto h^2$, giving that $A/A_0 = (h_c/h)^2$.}
\label{fig:Ah}
\end{figure}

\begin{figure}
  \centering
  \begin{overpic}[width=\linewidth]{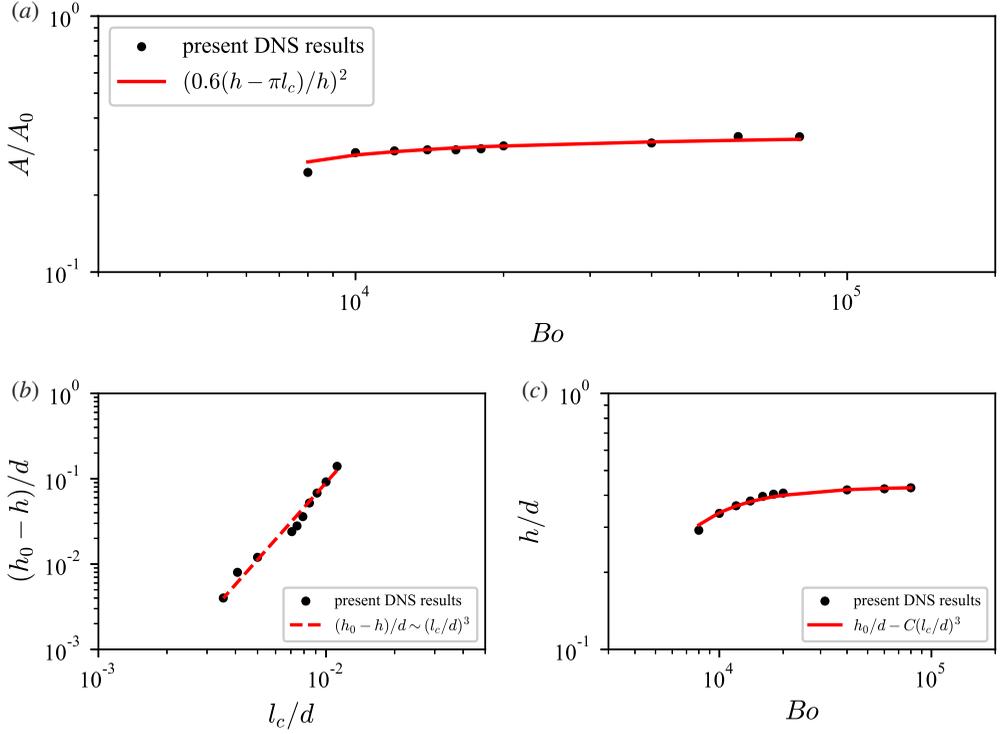}
  \put(2, 71){\small(\textit{a})}
  \put(2, 34){\small(\textit{b})}
  \put(52, 34){\small(\textit{c})}
  \end{overpic}
  \caption{Estimation of main cavity size and the breaking height. (a) cavity area over different Bond numbers. (b) Linear relationship between the decreased breaking height caused by shortened project distance and the capillary length, $(h_0-h)/d \propto ({l_c}/d)^3$. (c) A scaling to estimate the breaking height at different Bond numbers.}
\label{fig:Bo_A_h}
\end{figure}

The geometric properties of the main cavity caused by plunging breakers are identified by \citet{new1983class}, showing that the surface profiles underneath the overturning crest may be represented by an ellipse of axes ratio $\sqrt{3}$, with its major axis rotated at an angle of approximately $60^{\circ}$ to the horizontal. A similar shape can be confirmed in our cases as shown in figure \ref{fig:Ah}. The vertical height of the main cavity $h_c$, calculated as $h_c = h - h_t$, is closely related to the size of the main cavity entrapped by the plunging jet, where $h$ is the breaking height and $h_t$ is the height from the breaking crest to the cavity top. The cross-sectional area of the initially ingested cavity in the breaking process can be estimated by applying the ellipse area formula $A = \upi (h_c/sin60^{\circ})^2 / 4\sqrt{3}$, where $h_c$ is the vertical height of the main cavity. By normalizing the main cavity using the cross-sectional area $A_0$, we obtain $A/A_0 \propto (h_c/h)^2$. A new scaling regarding the cavity correction factor for the entrained cavity is proposed as $A/A_0 = ((h-\upi l_c)/h)^2$ by \citet{mostert2022high}, with very good agreement at high Bond numbers and weaker agreement at lower Bond numbers. This indicates that $h_c = h-\upi l_c$, where $l_c$ is the capillary length, defined as $\sqrt{d/Bo}$ in our study. Similar scaling can be proposed, but a coefficient of 0.6 should be used to mediate the difference between the width of the jet and the breaking height when it exhibits a greater separation between the wave scale and capillary length due to the larger Bond number in the present work, which gives $A/A_0 = (0.6(h - \upi lc)/h)^2$. Figure \ref{fig:Bo_A_h}(a) shows very good agreement between this scaling and the present DNS results.
As previously stated, the wave jet becomes thinner and projects further forward ahead of the wave as the surface tension decreases. It exhibits a breaking height $h_0$ in the absence of surface tension, which represents the maximum value of all breaking heights when surface tension is considered. It is found that the decreased breaking height caused by the shortened project distance normalized by $d$ is proportional to the cube of the capillary length normalized by $d$, which gives $(h_0-h)/d \propto ({l_c}/d)^3$, as shown in figure \ref{fig:Bo_A_h}(b), while $h_t$ remains constant under a distinct Bond number. Figure \ref{fig:Bo_A_h}(c) shows the comparison of the numerical results of $h/d$ to the estimated values of $h/d$ calculated using $h_0/d - C(l_c/d)^3$ by the proposed power-law scaling, with $C$ being a proportionality constant.

\subsection{Breaking criteria}\label{sec:Breaking criteria}
This section develops the relationship between wave parameters, i.e., maximum wave height before breaking $H$ [$\rm{L}$], breaking-wave crest $H_b$ [$\rm{L}$] of the plunging breaker, and the initial conditions used to generate waves in this study by numerical data fitting, following the above dimensional analysis as stated in (\ref{dimensionless parameters for wave characteristics}) and (\ref{dimensionless parameters of wave characteristics}):
\begin{equation}
\frac{H}{d} = f_H(\Rey, Bo, \frac{S}{d}, \frac{f}{\sqrt{g/d}})
  \label{dimensionless expression 1 for wave height}
\end{equation}
\begin{equation}
\frac{H_b}{d} = f_{H_b}(\Rey, Bo, \frac{S}{d}, \frac{f}{\sqrt{g/d}})
  \label{dimensionless expression 1 for breaking wave height}
\end{equation}
 As discussed in section \ref{sec:fluid proprieties}, $\Rey$ and $Bo$ do not significantly influence the wave characteristics, so the wave is considered to be independent of $\Rey$ and $Bo$ when discussing the scaling of $H$ and $H_b$ to the initial conditions. Thus,
\begin{equation}
\frac{H}{d} = f_H(\frac{S}{d}, \frac{f}{\sqrt{g/d}}) \propto (\frac{S}{d})^ {\alpha_H} (\frac{f}{\sqrt{g/d}})^ {\beta_H}
  \label{dimensionless expression 2 for wave height}
\end{equation}
\begin{equation}
\frac{H_b}{d} = f_{H_b}(\frac{S}{d}, \frac{f}{\sqrt{g/d}}) \propto (\frac{S}{d})^ {\alpha_{H_b}} (\frac{f}{\sqrt{g/d}})^ {\beta_{H_b}}
  \label{dimensionless expression 2 for breaking wave height}
\end{equation}
This dimensional analysis demonstrates the dependence of the wave characteristics on the dominant dimensionless variables derived from the initial conditions. Their quantitative relations are investigated by conducting various cases for different combinations of $S$, $f$, and $d$ to determine the corresponding coefficients in the dimensionless expressions.

First, the wave characteristics are estimated from the simplified theory for plane wavemakers. In shallow water, a simple theory for the generation of waves by wavemakers was proposed by Galvin (1964), who reasoned that the water displaced by the wavemaker should be equal to the crest volume of the propagating wave form. As breaking waves are generated by a piston wavemaker with a stroke of $S$ over a constant water depth $d$, the volume of water displaced over a whole stroke is $Sd$. If the resulting waves are vertically symmetric with one single crest before breaking, then the crest volume of the propagating wave forms in a wavelength is $\int_{0}^{L}(H/2)(1-\cos(2 \upi f x))\mathrm{d}x=HL/2$, where $L$ is the wavelength. Equating the two volumes,
\begin{equation}
Sd = \frac{HL}{2}
  \label{simple theory for wave (1)}
\end{equation}
According to the dispersion relation of shallow-water waves, the wavelength is $L = T \sqrt{gd}$. Then the resulting connection between the wave height and the initial conditions of the wave plate can be expressed as:
\begin{equation}
\frac{H}{d} = \frac{S}{T \sqrt{gd}}
  \label{simple theory for wave (2)}
\end{equation}
Notably, the wave parameters $H$, $L$, and $T$ in this expression are theoretical values and do not represent the real values in actual waves, which already break before forming a symmetrical waveform, but it provides us with a possible relationship that can be used to determine the fit to the numerical data.

Then, the scaling of the maximum wave height before breaking $H$ of the experimental waves generated by the wave plate is fitted through the numerical results under various initial conditions. It can be seen from equation (\ref{dimensionless expression 2 for wave height}) that $H/d \propto S^{\alpha_H} f^{\beta_H} d^{{\beta_H}/2 - \alpha_H} g^{-{\beta_H}/2}$, so $H \propto S^{\alpha_H} f^{\beta_H} d^{{\beta_H}/2 - \alpha_H + 1} g^{-{\beta_H}/2}$. At the same frequency $f$, numerical results show that $H/d \propto S d^{-1/2}$ and $H \propto S d^{1/2}$, so $\alpha_H / (\beta_H/2- \alpha_H) = -2$ and $\alpha_H / (\beta_H/2- \alpha_H + 1) = 2$, giving that $\alpha_H = 1$ and $\beta_H = 1$; thus, it gives:
\begin{equation}
\frac{H}{d} \propto \frac{Sf}{\sqrt{gd}} \propto \frac{U}{c}
  \label{relationship for wave height}
\end{equation}
where $U = S{\upi}f$ is the maximum wave plate velocity and $c = \sqrt{gd}$ is the linear velocity. This is quite similar to the theoretical result proposed in equation (\ref{simple theory for wave (2)}).
\begin{figure}
  \centerline{\includegraphics[width=0.7\linewidth]{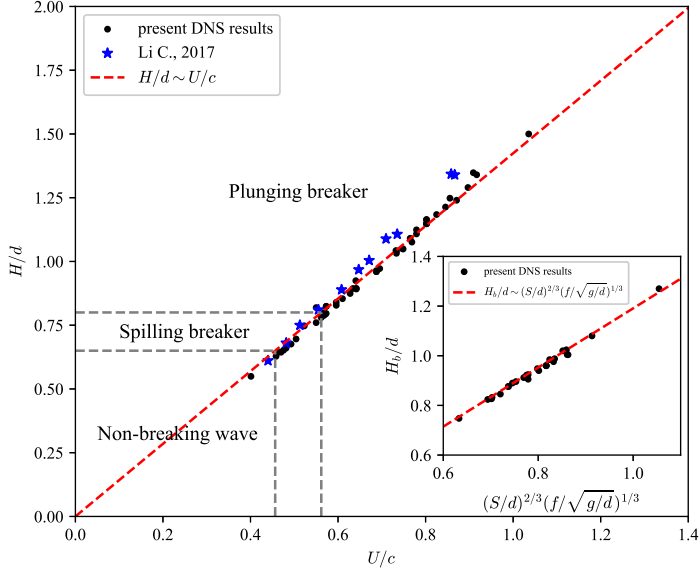}}
  \caption{Scaling for the maximum wave height before breaking, and breaking wave crest with respect to the initial conditions. Normalized wave height from equation (\ref{dimensionless expression 2 for wave height}) with the parameters of $\alpha_H = 1$ and $\beta_H = 1$. It indicates that the wave height normalized by the water depth is proportional to the maximum wave plate velocity normalized by wave phase speed. The inset shows the normalized breaking wave crest from equation (\ref{dimensionless expression 2 for breaking wave height}) with the parameters of $\alpha_{H_b} = 2/3$ and $\beta_{H_b} = 1/3$. Dash line: linear plot for comparisons.}
\label{fig:H-Vmax}
\end{figure}
Furthermore, the scaling of the breaking-wave crest $H_b$ with the initial conditions is also fitted through the numerical results. Based on the same method by analysing the numerical data, the exponents in power-law scaling can be determined as $\alpha_{H_b} = 2/3$ and $\beta_{H_b} = 1/3$, thus:
\begin{equation}
\frac{H_b}{d} \propto (\frac{S}{d})^ {2/3} (\frac{f}{\sqrt{g/d}})^ {1/3}
  \label{relationship for wave height at breaking}
\end{equation}
Figure \ref{fig:H-Vmax} shows the relationship between the normalized maximum wave height before breaking $H/d$ and breaking-wave crest $H_b/d$ to the initial conditions. A linear correlation between the maximum wave height before breaking $H$ and the maximum wave plate speed $U$ is revealed, showing that the wave height increases as the maximum wave speed increases. As indicated in figure \ref{fig:H-Vmax}, the generated wave remains nonbreaking for $H/d\le 0.65$. The breaking is of the spilling type for $0.65\ge H/d\le 0.80$, whereas it is of the plunging type for $H/d\ge 0.80$. The above results agree with the measurement performed by \citet{li2017dispersion}, who showed that the critical value for spilling and plunging waves is $H/d=0.80$. For plunging breakers, a linear correlation between breaking-wave crest $H_b$ and initial conditions is also proposed, which is in good agreement with the numerical results.

\subsection{Energy dissipation due to breaking}\label{sec:Energy dissipation}
The energy dissipation rate due to breaking can be defined as $\epsilon_l = \Delta E_m / \Delta t$, where $\Delta E_m$ is the decrease in the conservative energy $E_m = E_k + E_p$ during the active breaking period $\Delta t$. As shown in figure \ref{fig:wave123_E+e}(d), the energy dissipation appears to follow a linear trend with time but remains constant after approximately $\Delta t = 1/(2f)$, so in this study, the active breaking period starts when the wave breaks, and it has the duration of the movement of the wave plate $\Delta t = 1/(2f)$.

The physical parameters for the energy budget are the total energy  per unit length transferred by the motion of the wave plate $E_l$ [$\rm{ML/T^2}$], and the energy dissipation per unit length of the wave crest $\epsilon_l$ [$\rm{ML/T^3}$] for plunging breakers. Then, the dimensional analysis for the energy budget gives:
\begin{equation}
\frac{E_l}{\rho g d^3} = f_{E_l}(\Rey, Bo, \frac{S}{d}, \frac{f}{\sqrt{g/d}})
  \label{dimensionless expression 1 for E}
\end{equation}
\begin{equation}
\frac{\epsilon_l}{\rho g^{3/2} d^{5/2}} = f_{\epsilon_l}(\Rey, Bo, \frac{S}{d}, \frac{f}{\sqrt{g/d}})
  \label{dimensionless expression 1 for epsilon}
\end{equation}
Likewise, the energy budget is assumed to be independent of the Reynolds number and Bond number. Thus,
\begin{equation}
\frac{E_l}{\rho g d^3} = f_{E_l}(\frac{S}{d}, \frac{f}{\sqrt{g/d}}) \propto (\frac{S}{d})^{\alpha_{E_l}} (\frac{f}{\sqrt{g/d}})^{\beta_{E_l}}
  \label{dimensionless expression 2 for E}
\end{equation}
\begin{equation}
\frac{\epsilon_l}{\rho g^{3/2} d^{5/2}} = f_{\epsilon_l}(\frac{S}{d}, \frac{f}{\sqrt{g/d}}) \propto (\frac{S}{d})^{\alpha_{\epsilon_l}} (\frac{f}{\sqrt{g/d}})^{\beta_{\epsilon_l}}
  \label{dimensionless expression 2 for epsilon}
\end{equation}
The scaling is connected to an inertial model of the energy dissipation rate for our breaking waves, which follows the inertial estimate used by \citet{drazen2008inertial} in deep-water breakers and \citet{mostert2020inertial} for shallow-water breakers. 
To relate this isotropic turbulence assumption to the empirical relationship fitted from the numerical results, a reasonable approximation of the turbulent integral length scale $l$, the characteristic velocity scale $w$ and the turbulent cloud cross-section $A$ should be derived from the breaking amplitude $H_b$ and water depth $d$.

During the postbreaking period in our breaking waves, a triangular area with highly developed turbulence is generated at the upper surface of the water column. In the process of deep-water breaking waves, this turbulent region, tracked by photographing the evolution of a dye patch in the laboratory measurements of \citet{rapp1990laboratory}, reaches depths of two to three wave heights and horizontal lengths of approximately one wavelength within five wave periods of breaking. As shown in the vorticity field during the postbreaking period in figure \ref{fig:wave1_colormap_b}(i), the turbulence cloud at the upper surface is considered to be a triangular area that extends over a horizontal length of one wavelength $L$ and a vertical height of one breaking-wave height $H_b$ since $H_b$ describes an effective diameter for the amount of available water supplied to the turbulent cloud. Therefore, the turbulence cloud cross section $A$ is assumed to be proportional to $L{H_b}/2$. The representative velocity scale $w$ and turbulent integral length scale $l$ can be regarded as the phase speed $\sqrt{gd}$ and wavelength $L$, respectively. Thus the dissipation per unit length along the wave crest is:
\begin{figure}
  \centering
  \begin{overpic}[width=0.5\linewidth]{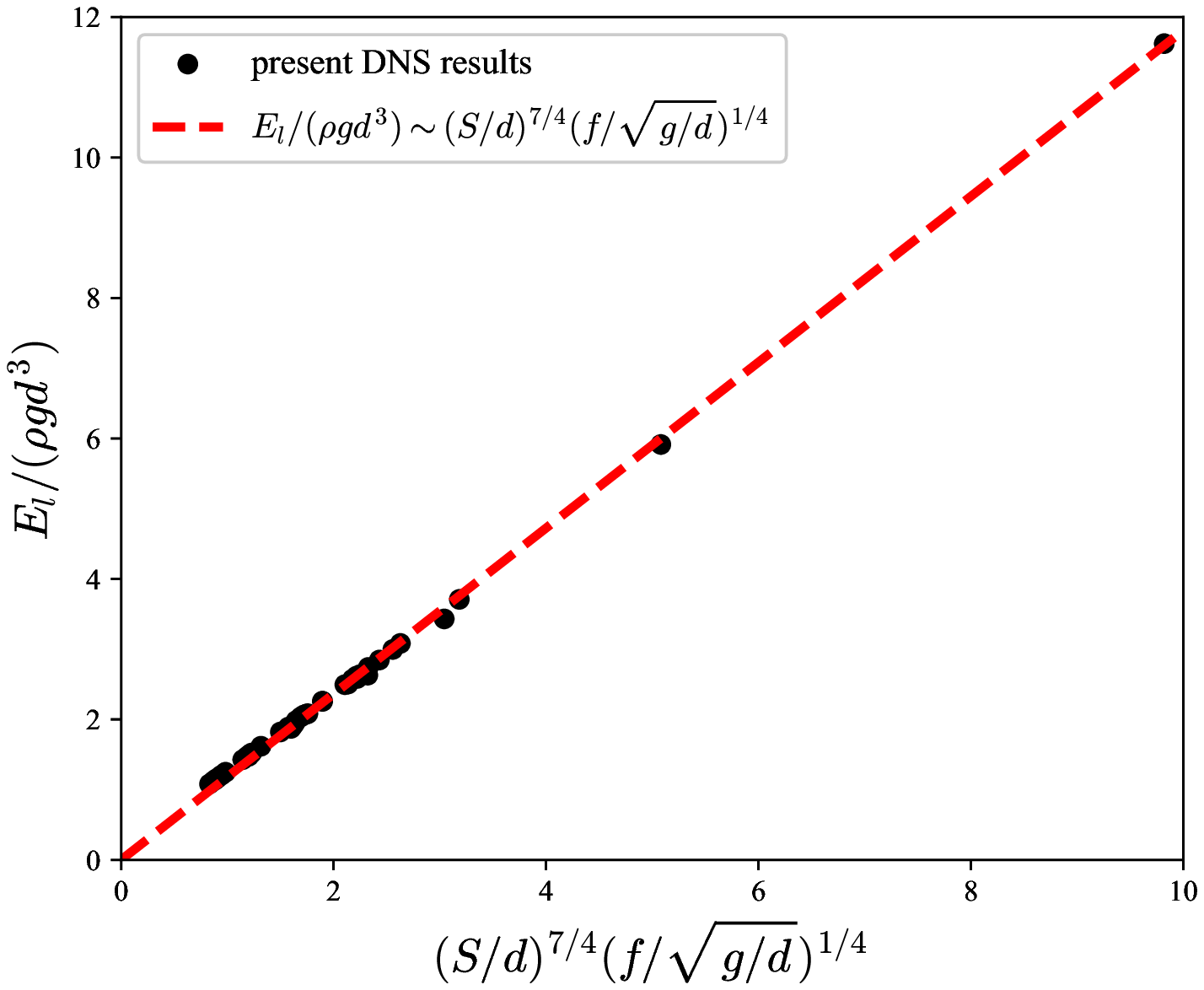}
  \put(0,78){\small(\textit{a})}
  \end{overpic}
  ~
  \begin{overpic}[width=\linewidth]{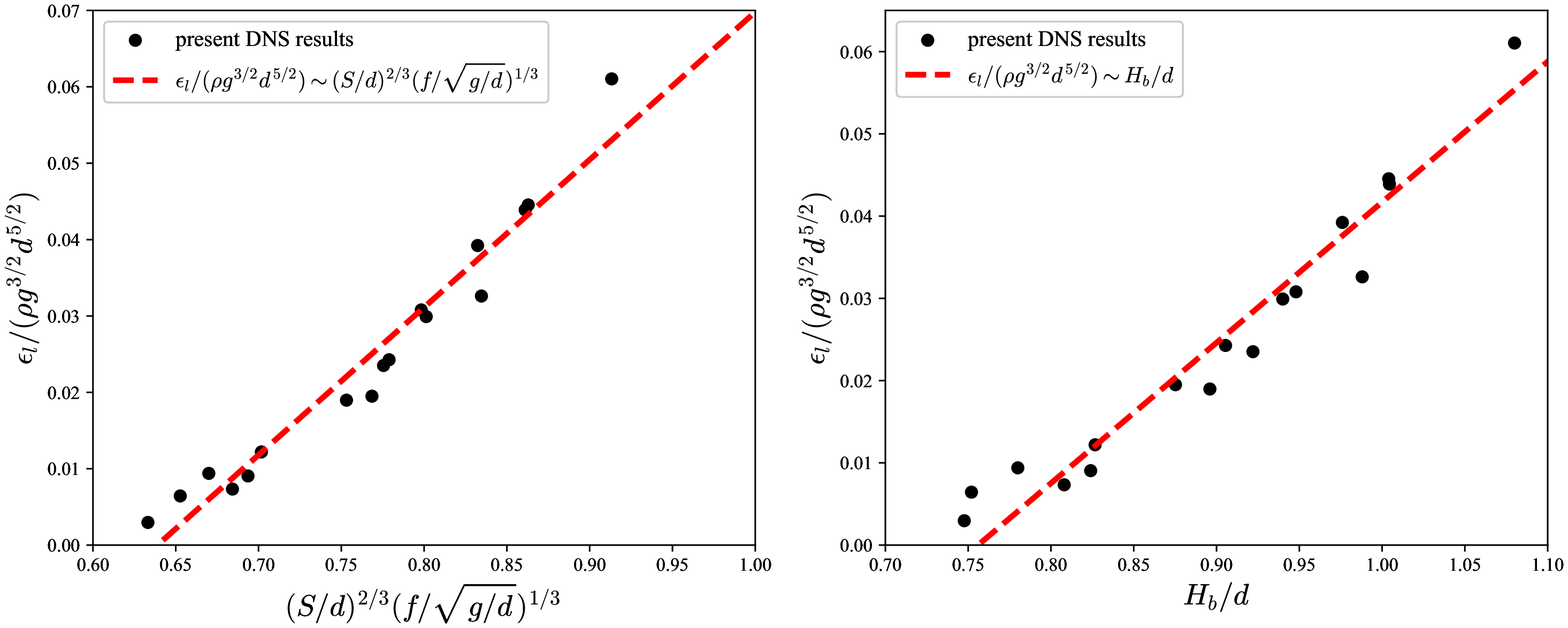}
  \put(0,38){\small(\textit{b})}
  \put(50,38){\small(\textit{c})}
  \end{overpic}
  \caption{Scaling for the total energy transferred by the motion of wave plate $E_l$ (a), and the energy dissipation per unit length of breaking wave $\epsilon_l$ (b) with respect to the initial conditions. (a) Normalized total energy from equation (\ref{dimensionless expression for E}) with the parameters of $\alpha_{E_l} = 7/4$ and $\beta_{E_l} = 1/4$. (b) Normalized energy dissipation rate from equation (\ref{dimensionless expression for epsilon}) with the parameters of $\alpha_{\epsilon_l} = 2/3$ and $\beta_{\epsilon_l} = 1/3$. Dash line: linear plot for comparisons. (c) Scaling for energy dissipation per unit length of breaking wave $\epsilon_l$ with respect to local breaking parameters $H_b/d$ as shown in equation (\ref{energy dissipation rate2}).}
\label{fig:Eee}
\end{figure}
\begin{equation}
\epsilon_l = \rho_w A \epsilon \propto \rho_w \frac{H_bL}{2} \frac{\sqrt{gd}^3}{L} \propto \rho_w g^{3/2} H_b d^{3/2}
  \label{simple theory for energy dissipation rate 1}
\end{equation}
\begin{equation}
\frac{\epsilon_l}{\rho g^{3/2} d^{5/2}} \propto \frac{H_b}{d}
  \label{simple theory for energy dissipation rate 2}
\end{equation}
Next, the scaling of the energy budget to the initial conditions is determined by numerical data fitting. When $\alpha_{E_l} = 7/4$, $\beta_{E_l} = 1/4$, and $\alpha_{\epsilon_l} = 2/3$, $\beta_{\epsilon_l} = 1/3$, the best linear relations can be obtained by least square fitting. The relationships between the total energy transferred by the motion of the wave plate $E_l$, the energy dissipation per unit length of breaking wave $\epsilon_l$, and the initial conditions can be described as:
\begin{equation}
\frac{E_l}{\rho g d^3} \propto (\frac{S}{d})^{7/4} (\frac{f}{\sqrt{g/d}})^{1/4}
  \label{dimensionless expression for E}
\end{equation}
\begin{equation}
\frac{\epsilon_l}{\rho g^{3/2} d^{5/2}} \propto (\frac{S}{d})^{2/3} (\frac{f}{\sqrt{g/d}})^{1/3}
  \label{dimensionless expression for epsilon}
\end{equation}
As stated in section \ref{sec:Breaking criteria}, based on the relationship of the normalized breaking-wave crest to the initial conditions $H_b/d \propto (S/d)^{2/3}(f / \sqrt{g/d})^{1/3}$, the energy dissipation rate can be connected to the local breaking parameters:
\begin{equation}
\frac{\epsilon_l}{\rho g^{3/2} d^{5/2}} \propto (\frac{S}{d})^{2/3} (\frac{f}{\sqrt{g/d}})^{1/3} \propto \frac{{H_b}}{d}
  \label{energy dissipation rate1}
\end{equation}
This link between the energy dissipation rate and the local breaking parameters is applicable to plunging breakers. As discussed in section \ref{sec:Breaking criteria}, the generated wave breaks as a plunging type at $H/d \ge 0.80$. This shows that no energy is dissipated by plunging breakers when $H/d < 0.80$, which corresponds to $H_b/d < 0.76$ according to the relationship between $H$ and $H_b$. This indicates the following:
\begin{equation}
\frac{\epsilon_l}{\rho g^{3/2} d^{5/2}} \propto \frac{{H_b}}{d} = \chi(\frac{H_b}{d} - 0.76)
  \label{energy dissipation rate2}
\end{equation}
where $\chi$ is a proportionality constant.
Figure \ref{fig:Eee} shows a good linear dependence between the energy dissipation per unit length of breaking wave and the local breaking parameters. This scaling demonstrates that the dominant dimensionless variable describing the energy dissipation by breaking of the experimental waves generated by the wave plate is the ratio of $H_b$ and $d$.


\section{Concluding remarks}\label{sec:Concluding remarks}
The present study was designed to determine the effect of the fluid properties and initial conditions on the dynamics, kinematics, energy dissipation, and air entrainment in the breaking process, by performing high-fidelity simulations of breaking waves generated by a piston-type wave plate using direct numerical simulation. The investigation of the stroke and frequency of the wave plate has shown detailed information, including breaking characteristics, energy transfer and dissipation, and air entrainment during wave breaking. A quantitative relationship between the main cavity size and the breaking height is presented based on the investigation of the influence of the Bond number on the evolution of the overturning jet. This reveals the effect of surface tension on the crest overturning process, which thickens the width of the plunging jet and shortens the distance that projects forward ahead of the wave. The resulting wave height is estimated based on the simplified theory for plane wavemakers, and a reliable agreement is obtained between this theoretical result and our numerical data. The link between wave height and initial conditions indicates that waves can be classified as nonbreaking waves, spilling breakers, and plunging breakers based on the ratio of wave height to water depth $H/d$. A scaling of the breaking-wave crest to the initial conditions for the plunging breaker is also presented, showing a clear linear dependence. Additionally, the conventional dissipation scaling of turbulence theory is applied to the wave-breaking process, deriving a link between the energy dissipation rate and the ratio of breaking-wave crest to the water depth $H_b/d$, which is supported by our numerical data. The proposed scaling laws quantitatively link the kinematics and dynamics of breaking waves to the initial conditions, which may be of use for future theoretical analysis.

\backsection[Acknowledgements]{We appreciate beneficial discussions and help from the BASILISK community. Simulations were performed using computational resources on Advanced Research Computing (ARC) at Virginia Tech.}

\backsection[Funding]{This work has been supported by the scholarship from China Scholarship Council (CSC) under the Grant No. 201906090270.}

\backsection[Declaration of interests]{The authors report no conflict of interest.}

\backsection[Author ORCIDs]{S. Liu, https://orcid.org/0000-0002-8530-8359; H. Wang, https://orcid.org/0000-0001-9733-0150.}

\bibliographystyle{jfm}
\bibliography{references}

\end{document}